\documentclass[twocolumn,aps,floats,prd,psfig,amsmath,amssymb]{revtex4}
\usepackage{graphicx, epsfig, natbib}
\usepackage{subfigure}
\usepackage{color}

\newcommand{\quadmod}{Eq.~(\ref{set_max})}

\newcommand{\wj}[6]{\left(
                           \begin{array}{ccc}
        \! #1\! & #2\!  & #3\!  \\
        \! #4\! & #5\!  & #6\!
                           \end{array}
                   \right)}

\begin{document}

\title{Testable polarization predictions for models of CMB isotropy anomalies}
	\author{} 
\author{Cora Dvorkin,$^{1, 2}$ Hiranya V. Peiris,$^{1, 3}$\footnote{Hubble Fellow} and Wayne Hu$^{1}$}
\email{cdvorkin@uchicago.edu} 

\affiliation{{}$^1$Kavli Institute for Cosmological Physics, Department of Astronomy \& Astrophysics, and Enrico Fermi Institute, University of Chicago, Chicago IL 60637, U.S.A. \\
{}$^2$Department of Physics, 
University of Chicago, Chicago IL 60637, U.S.A.\\
{}$^3$Institute of Astronomy, University of Cambridge, Cambridge CB3 0HA, U.K.
}

\date{\today}
\begin{abstract}
Anomalies in the large-scale CMB temperature sky measured by WMAP have been suggested as possible evidence for a violation of statistical isotropy on large scales.  In any physical model for broken isotropy, there are testable consequences for
the CMB polarization field.   We develop simulation tools for predicting the polarization field
in models that break statistical isotropy locally through a modulation field.  
We study two different models: dipolar modulation,  invoked to explain 
the asymmetry in power  between northern and southern ecliptic hemispheres, and 
quadrupolar modulation, posited to explain the alignments between the quadrupole and octopole.
For the dipolar case, we show that predictions for the correlation between the first 10 multipoles 
of the temperature and polarization fields can typically be tested at better than the 98\% CL.  For the quadrupolar case, we show that the polarization quadrupole and octopole should be
moderately aligned. Such an alignment is a generic prediction of explanations which
involve the temperature field at recombination and thus discriminate against explanations involving foregrounds or local secondary anisotropy.  Predicted
correlations between temperature and polarization multipoles out to $\ell = 5$
provide tests at the $\sim$ 99\% CL or stronger for quadrupolar models that make the temperature alignment
more than a few percent likely.  
As predictions of 
anomaly models, polarization statistics move beyond the {\it a posteriori} inferences
that currently dominate the field.
\end{abstract}

\maketitle
\section{Introduction}

Several anomalies observed in the CMB data suggest a possible breaking of statistical isotropy on the largest angular scales. 
Broken isotropy would represent a radical revision of the standard cosmological model
as statistical isotropy underlies  all cosmological inferences.  As such, it is important
to test such claims by all means possible.  

Specifically, 
the measurements from the {\it Wilkinson Microwave Anisotropy Probe} (WMAP) \cite{bennett, spergel, wmap3yr, page} indicate an asymmetric distribution in power between the northern and southern ecliptic hemispheres \cite{eriksen_hansen}. 
Also, the data exhibit alignments in the temperature field; in particular, the observed quadrupole and octopole of the CMB are planar and mutually aligned, and this plane is oriented roughly orthogonal to the CMB dipole direction
\cite{oliveira,land_and_magueijo}.

The statistical significance of these preferred directions
has been investigated by several authors \cite{oliveira, tegmark, copi,schwarz,axis, eriksen_banday,Copi:2005ff,Copi:2006tu,Wiaux:2006zh,Land:2004bs}, 
but the {\it a posteriori} choice of statistics makes their interpretation difficult
\cite{Magueijo:2006we,Land:2006bn,Gordon:2007xm}.   Indeed several 
generic tests of isotropy yield no evidence for statistical anisotropy 
\cite{Hajian:2003qq,Hajian:2006ud,Donoghue:2004gu,ArmendarizPicon:2005jh}.
Since measurements of the temperature field have already reached the cosmic variance
limit, further information from it will not be forthcoming, barring 
improvements in foreground and systematics modeling.
 On the other hand, large-scale CMB polarization measurements are still in their infancy and provide fertile grounds for future tests of
isotropy.  While generic tests of isotropy have also been proposed for polarization 
\cite{Basak:2006ew,Pullen:2007tu} and unrelated statistical anomalies may be found, 
any physical model that purports to explain the temperature anomalies
provides testable predictions for the statistics of the polarization field.   Matching 
anomalies in polarization for such models
can provide a means of going beyond both {\it a posteriori} inferences and blind
searches for isotropy anomalies.

In this {\it Paper}, we develop the radiative transfer tools required to 
investigate polarization statistics.  Large-angle polarization arises from rescattering of CMB radiation
from recombination during reionization.  Given a model that predicts the statistical
properties of the temperature field
at recombination, Monte-Carlo realizations of the polarization field can be generated
and analyzed for signatures of broken isotropy.

We apply these techniques to models that break statistical isotropy with a multiplicative
modulation field \cite{wayne}.  These models posit that the observed temperature field
is the product of a modulating field with superhorizon wavelength fluctuations and a second field
that carries fluctuations at the observed scales.   Gradients in the modulation field naturally
pick out a preferred direction within our horizon.
Such models have been proposed to explain both the hemisphere asymmetry 
\cite{spergel,gordon,eriksen_banday} and the quadrupole-octopole alignment \cite{wayne}.

The structure of the paper is as follows.  In \S~\ref{sec: isotropy_breaking} we discuss the general form of the modulation model and develop the simulation tools to predict the
polarization field.  
The analytic framework for polarization predictions and explicit relations for
the simplest modulation statistics are given in the Appendix.
In \S~\ref{sec: dipolar}, we study the polarization predictions for the dipolar modulation models
that are designed to explain the hemisphere asymmetry.  
We continue in  \S~\ref{sec: quadrupolar} by studying quadrupolar modulation models that
explain the quadrupole-octopole alignment.  We discuss our results in \S~\ref{sec: discussion}.


\section{Spontaneous Isotropy Breaking} \label{sec: isotropy_breaking}

\subsection{Modulation Model} \label{sec: general_model}

We consider a model where a superhorizon scale modulation of the gravitational potential field causes the CMB temperature field to locally look statistically anisotropic, even though globally the model preserves statistical homogeneity and isotropy.   
We generalize the considerations of Gordon {\it et al.}~\cite{wayne} by explicitly working with a 3D model
of the gravitational potential as required for a polarization analysis.

Let us begin by assuming that the source of  gravitational potential or curvature fluctuations depends on the product of two fields $g({\bf x})$ and $h({\bf x})$.  For example, these curvature fluctuations might arise from an inflaton decay rate that depended on two fields, each with their own quantum fluctuations.
Let us further assume that the field $h({\bf x})$ has only superhorizon scale fluctuations so that in any given Hubble volume, it takes on a deterministic value whereas $g({\bf x})$ has subhorizon scale power that appears as stochastic fluctuations within the volume. 

Within our Hubble volume, an observer would see broken statistical homogeneity from the slow modulation of $h({\bf x})$ across the volume.  Furthermore, the local gradient and curvature of $h({\bf x})$ naturally picks out a direction and breaks statistical isotropy as well.  
This broken isotropy is then transferred to the CMB through the Sachs-Wolfe effect \cite{sachs_wolfe} for the temperature field and Thomson scattering during reionization for the polarization field.  Nonetheless, a full spatial average over many Hubble volumes would show no statistical inhomogeneity or anisotropy.

 To make these considerations compatible with the observed statistical homogeneity and isotropy on small scales, we add another field that is responsible for fluctuations well below the horizon scale.
 Specifically, we model the Newtonian curvature fluctuation as
\begin{equation} \label{general_model1}
\Phi ({\bf x})=  g_1 ({\bf x})\left[1+h({\bf x})\right] + g_2 ({\bf x})\,,
\end{equation}
where $g_1$($\bf{x}$) and $g_2$($\bf{x}$) are Gaussian random fields 
and $h({\bf x})$ is the modulating long-wavelength
field. 
We will refer to the latter as the ``modulating field'' throughout this work.

Several important properties of this model are revealed in Fourier space.    The product of
fields in physical space becomes a convolution in Fourier space
\begin{eqnarray} \label{convolution}
\Phi({\bf k}) = g_1({\bf k})  +  \int \frac{d^{3}k^{\prime}}{(2 \pi)^{3}}g_{1}({\bf k}^{\prime}) 
h({\bf k}-{\bf k}^{\prime})
+ g_{2}({\bf k})\,. 
\end{eqnarray}
Two point correlations in $g_1$ and $g_{2}$ are determined by their respective power spectra
\begin{eqnarray}
\langle g^{*}_1({\bf k})g_1 ({\bf k}^\prime)\rangle &=& (2\pi)^3 \delta({\bf k}-{\bf k'}) P_{g_1}(k) \,,\nonumber\\
\langle g^{*}_2 ({\bf k})g_2 ({\bf k}^\prime)\rangle &=& (2\pi)^3 \delta({\bf k}-{\bf k'}) P_{g_2}(k) \,.
\end{eqnarray}
We will typically assume that $g_1$ has power in fluctuations near the horizon scale
and $g_2$ has power on subhorizon scales.  The modulating field $h$ has only superhorizon
modes.  
That the two point correlations do not couple modes of different ${\bf k}$ is a consequence
of statistical homogeneity and that the power spectrum is a function of $k = | {\bf k}|$ only
is a consequence of statistical isotropy.

Notice that in the presence of the modulation $h$, $\Phi$ does not obey statistical homogeneity
and isotropy
\begin{eqnarray}
\label{phi2pt}
\langle\Phi^*({\bf k})\Phi({\bf k}^\prime)\rangle
&=& (2\pi)^3 \delta({\bf k}-{\bf k'}) [  P_{g_1}(k) + P_{g_2}(k) ] \nonumber\\
&& +  [ P_{g_1}(k) +P_{g_1} (k') ]  h({\bf k}^{\prime}-{\bf k})\\ 
&& +  \int {d^3 \tilde k \over (2\pi)^3} P_{g_1}(\tilde k) 
 h^{*}({\bf k}-\tilde{\bf k}) h({\bf k}^\prime -\tilde{\bf k})\,, \nonumber
\end{eqnarray}
where the ensemble average is over realizations of $g_1$ and $g_2$ only.
In particular, ${\bf k}$ modes that are separated by less than
the horizon wavenumber are correlated and their 
amplitude depends on the direction of ${\bf k}$ despite the statistical isotropy of $g_1$.

\subsection{Temperature Field}

Broken homogeneity and isotropy are transferred onto the CMB temperature field
through the Sachs-Wolfe
effect.   Horizon scale fluctuations in the Newtonian curvature at recombination imprint 
large scale temperature perturbations as 
\begin{equation}\label{sachs-wolfe}
{\Delta T \over T}({\bf x}) = -{1\over 3}\Phi({\bf x})\,.
\end{equation}
To the observer at the origin,  temperature inhomogeneities on a shell 
at the distance $D_{\rm rec}$ that
a photon travels since recombination appear as anisotropies in a direction $\hat {\bf n}$
\begin{equation}
{\Delta T \over T}(\hat {\bf n}) = {\Delta T \over T}({\bf x}=D_{\rm rec} \hat {\bf n})\,.
\end{equation}
The analogue of a Fourier decomposition for the potential field is the spherical harmonic
decomposition of the temperature field
\begin{equation}
{\Delta T \over T}(\hat{\bf n})=\sum_{\ell}\sum_{m=-\ell}^{\ell}T_{\ell m}Y_{\ell m}(\hat{\bf n})\,,
\end{equation}
where $Y_{\ell m}$ are  spherical harmonics.

Given a statistically homogeneous and isotropic potential field, the two point correlations in the
temperature field obey
\begin{equation}
\langle T_{\ell^\prime m^{\prime}}^{*}T_{\ell m} \rangle=C_{\ell}^{TT}\delta_{\ell \ell^{\prime}}\delta_{m m^{\prime}}\,,
\end{equation}
where $C_{\ell}^{TT}$ is the angular power spectrum of the temperature field.  In the presence
of the modulating field, the properties of the two point correlation of the potential in 
Eq.~(\ref{phi2pt}) carry over to the temperature field: modes of neighboring $\ell$ are correlated
and the power depends on direction, {\it i.e.}\ the value of $m$.  

\begin{figure}[tb]
\begin{center}
\includegraphics[width=2.75in]{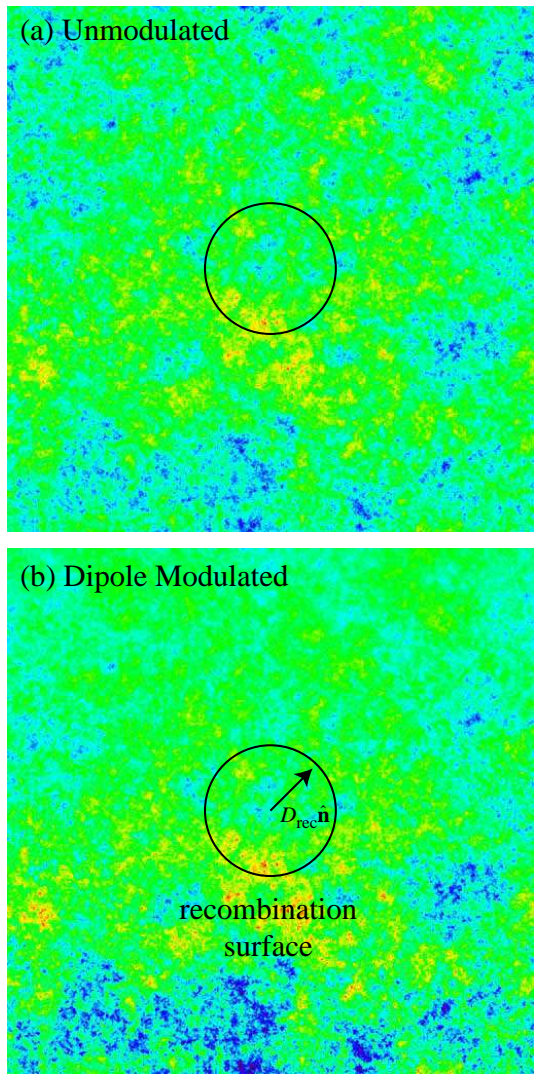}
\caption{Simulations of modulated temperature field.  (a) Gaussian random unmodulated field $g_{1}$.  (b) Dipole modulated potential field with $h({\bf x}) = w_{1} 
\sqrt{3/4\pi}(z/D_{{\rm rec}})$
where the $z$ axis is the modulation axis.
   Also shown (circle) is the surface at recombination upon
which the observed temperature field originates.  Note the power asymmetry in the
modulated field.} \label{sim}
\end{center}
\end{figure}

The two point  correlation of the model in terms of the power spectra of the fields $g$ and $h$ 
was given in \cite{wayne} and is rederived in our framework of 3D fields in \S \ref{sec: modulated_temp_statistics}.     In this framework, the modulation appears as the product of the Gaussian random field $g_{1}({\bf x})$ and the modulation field $h({\bf x})$ to form the observed temperature field.  For the temperature field, this is equivalent to the product of the 2D angular fields projected onto the recombination surface (see Fig.~\ref{sim}, solid circle). 
Likewise the two point correlation can be expressed in terms of operations on the
angular power spectra of the $g$ and $h$ fields in projection.   
 We shall now see that
these simplifications do not carry over to the polarization field.

\subsection{Polarization Field} \label{sec: polarization}

Broken homogeneity and isotropy are also transferred onto the CMB polarization field.  
A modulation model that is designed to explain temperature anomalies can be tested
through the predictions it makes for polarization.  

\begin{figure}[tb]
\begin{center}
\includegraphics[width=3in]{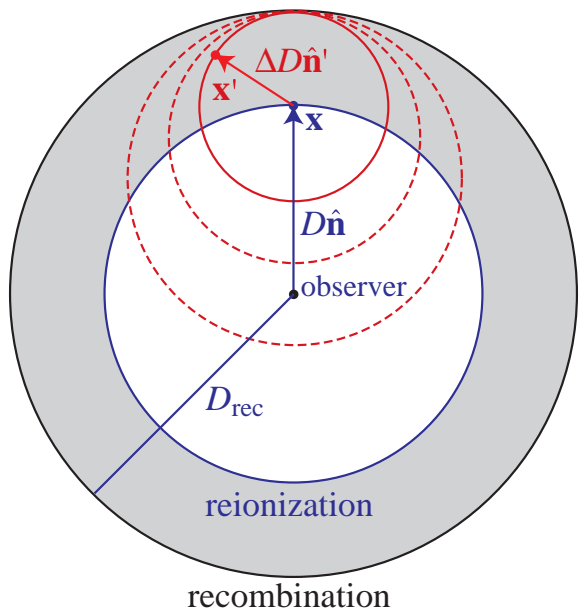}
\caption{Schematic diagram for polarization generation.  After reionization, free electrons along the
line of sight $\hat {\bf n}$ at a position ${\bf x}$ see a quadrupole anisotropy on their
recombination surfaces at a distance $\Delta D$ in their angular coordinates
${\hat {\bf n}}'$.   Thomson scattering off these electrons produces the polarization field
viewed by the observer at the origin in direction $\hat {\bf n}$.} \label{pol}
\end{center}
\end{figure}

Large-angle polarization is generated by the rescattering of photons off free electrons
 during the reionization epoch.    A quadrupole anisotropy in the temperature of the
 radiation leads to linear polarization of the CMB.   More specifically, from the temperature
 field at recombination of Eq.~(\ref{sachs-wolfe}), the quadrupole moments at each
 position in space are
given by
\begin{equation}
\label{quadsource}
T_{2m}({\bf x}) = \int d\hat{\bf{n}}^\prime Y_{2m}^{*}(\hat{\bf{n}}^\prime ){\Delta T \over T}
(
{\bf x}^\prime)\,.
\end{equation}
Note that $T_{2m}$ describes a set of 5 three dimensional source fields.  
Furthermore the angular integral is over the
recombination surface  of an (electron) observer at 
position ${\bf x}$ a distance $D$ away from the true observer at the origin
\begin{equation}
\label{coordinates}
{\bf x}' =  {\bf x} + \Delta D \hat{\bf{n}}^\prime \,,
\end{equation}
where $\Delta D = D_{\rm rec}-D$ is the radius of recombination surface of the electron, as illustrated by Fig.~\ref{pol}.

To the true observer at the origin, scattering at position ${\bf x}=D\hat{\bf n}$ generates a contribution to
the Stokes parameters of the polarization in direction $\hat {\bf n}$ given by \cite{reionization_rev}
\begin{eqnarray} \label{qplusiu1}
(Q \pm i U)(\hat{{\bf n}})&=&-\frac{\sqrt{6}}{10} \int dD {d\tau \over dD} e^{-\tau(D)}
 \nonumber\\
&&\times \sum_{m=-2}^{2}T_{2m}(D  \hat{\bf{n}} ){}_{\pm2}Y_{2m}(\hat{\bf{n}})\,,
\end{eqnarray}
where ${}_{\pm2}Y_{\ell m}$ are the spin $\pm 2$ spherical harmonics
and $\tau(D)$ is the optical depth between a comoving distance $D$ and the present epoch.
When we specify $\tau$ without the argument $D$, we refer to the total 
optical depth to reionization.  
The analogue of temperature multipole moments $T_{\ell m}$
 for linear polarization are the $E_{\ell m}$ and $B_{\ell m}$
multipole moments
\begin{equation}
\label{qplusiumain}
(Q \pm i U)(\hat{\bf{n}})=-\sum_{\ell m}\left[E_{\ell m} \pm i B_{\ell m}\right] {}_{\pm2}Y_{\ell m}(\hat{\bf{n}})\,.
\end{equation}
$E$ and $B$ modes are the tensor analogues of curl-free and divergence free components
of a vector.   An $E$-mode has a polarization direction that is aligned with or orthogonal to
the direction that the mode amplitude changes.  A $B$-mode has this direction rotated by
$\pm 45^\circ$.   When polarization is generated by a scalar field, only $E$-modes are formed.  
In Fourier space, this property is easily seen since the polarization direction from each ${\bf k}$ mode
must follow the direction of $\hat{\bf k}$ itself.
Since the modulation preserves the  scalar nature of the source of polarization, $B$-modes
are absent even in the modulated case.   Note that this is different from a direct modulation of the polarization
amplitude itself by a scalar quantity, e.g. from spatial fluctuations in $\tau$ \cite{reionization_rev}.   
In that case, the polarization direction 
is fixed by the potential field from recombination but the amplitude carries additional directionality
from the modulation.

Predictions  for the modulated polarization field, unlike the temperature field, require a 3D model
of the modulation.   Eq.~(\ref{coordinates}) implies that the relevant modulation is 
at position ${\bf x}'$
in a sphere around each source at position ${\bf x}=D\hat{\bf n}$ for each 
angular position $\hat{\bf n}$ (see  Fig.~\ref{pol}). 
Furthermore, these sources lie in the interior of the recombination surface
$D < D_{\rm rec}$ of the temperature
field.   Likewise the polarization statistics cannot be expressed as an operation on the
angular power spectra of the fundamental $g$ and $h$ fields corresponding to a single
projection.   The analytic framework for predicting the two point correlations is given in
\S \ref{sec :general_polarization}.  Given the cumbersome nature of this  framework,
we instead rely on numerical simulations to study modulated statistics as we shall now describe.
We use the analytic framework to test these simulations (see \S \ref{sec: Codetests})
and to provide insight on statistical quantities that should be measured in the 
simulations and data (see \S \ref{sec: Unbiased estimator}).

\subsection{Simulations} \label{sec: simulations}

Numerical simulations are conceptually the simplest and most general way of assessing the
statistics of the modulated temperature and polarization fields.   

We first simulate the 3D potential field $\Phi({\bf x})$ by combining realizations of the two Gaussian
random $g$-fields with the deterministic modulation field $h$.
 We typically chose the length of the periodic box to be four times the diameter of the 
 recombination surface with $512^3$ pixels. This ratio of sizes and dynamic range allows us to cover the $2 \le \ell \lesssim 10$ scales of interest in this work.
 
 To obtain the temperature field,
we then interpolate the field onto a 2D surface of radius $D_{\rm rec}$ and weight it according to the Sachs-Wolfe relation (\ref{sachs-wolfe}). 
Finally we use the HEALPix package \cite{healpix} to extract the multipole moments of the field.  The interpolation is done in every unit vector $\hat{\bf {n}}$ corresponding to $12,288$ HEALPIX pixel centers at resolution $N_{\rm{side}}=32$.
 As an example, slices through simulations of unmodulated and dipole-modulated temperature fields are shown in Fig.~\ref{sim}.   The statistics of these temperature simulations match
 those of the 2D approach where the modulation is directly in angular space on the
 recombination surface.  We will occasionally use 2D simulations where only the
 temperature field is required and for unmodulated cases to build large samples.

The polarization field is equally straightforward to simulate but numerically more involved.
We take the same 3D simulations employed for the temperature field and compute the quadrupolar temperature sources for $80$ redshifts in equal steps along each line of sight, and in every unit vector $\hat{\bf {n}}$ corresponding to $12,288$ HEALPIX pixel centers at resolution $N_{\rm{side}}=32$ around the sky.  Then, we generate maps of polarization by means of Eq.~(\ref{qplusiu1}).
Once the maps are created, the coefficients $E_{\ell m}$ are obtained using the HEALPix package \cite{healpix}, and various polarization statistics described below are computed.

We test this pipeline against analytic calculations of unmodulated and 
simple modulation models in
\S \ref{sec: Codetests}.  
It is the main tool used to assess the temperature and polarization statistics in this work.
It can be used for arbitrarily complicated modulation fields and any statistical property of the
temperature and polarization fields.

\section{Dipolar Modulation}
\label{sec: dipolar}

Eriksen {\it et al.}~\cite{eriksen_hansen} showed that there is a hemispherical or dipole asymmetry in the amount
of power in the WMAP data at multipoles $\ell\sim 20-40$.    The axis of this asymmetry 
points in a direction that is close to the South Ecliptic Pole (SEP) making the power in the southern
hemisphere larger than the north.  It was subsequently pointed out that a dipolar modulation of the
form of Eq.~(\ref{general_model}) with a modulation field of
\begin{eqnarray}\label{dipole_parameters}
h(x) &=& w_{1}Y_{10}
\left({z \over D_{\rm {rec}}}\right)\,,
\end{eqnarray}
could explain the asymmetry though not the curious alignment with the SEP 
\cite{spergel,gordon,eriksen_banday}.  
Note that $Y_{10}(z/D_{{\rm rec}})= \sqrt{3/4\pi}(z/D_{{\rm rec}})$ in a coordinate 
system where the $z$-axis points along the modulation axis.
We set the $g_2$ field to have power only at $\ell > 40$ and hence its contribution is
irrelevant in our considerations.  
We take the $g_1$ field to have a scale invariant power spectrum
\begin{eqnarray}
 \frac{k^{3} P_{g_1}(k)}{2 \pi^{2}}  = {9 \over 25} \delta_{\zeta}^{2}
 \label{LCDMmodel}
\end{eqnarray}
with $\delta_{\zeta}=4.6 \times 10^{-5}$ which corresponds to
the WMAP 3-year normalization for a cosmology with 
 matter density $\Omega_{m}=0.24$, cosmological constant $\Omega_{\Lambda}=0.76$, baryon density $\Omega_{b}h^{2} = 0.022$, Hubble constant $H_{0}=73$ km s$^{-1}$ Mpc$^{-1}$, and spatially flat spatial geometry.   
 
\begin{figure}[htbp]
\begin{center}
\includegraphics[width=3.3in]{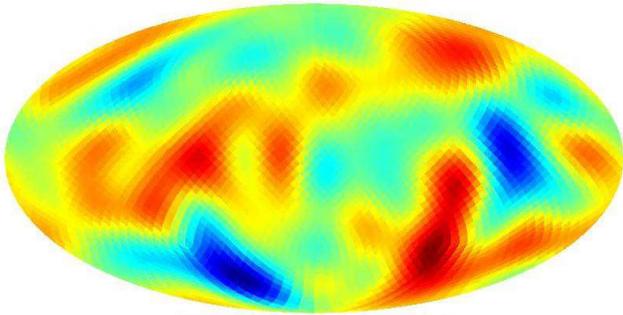}
\caption{Map composed of the first $10$ multipoles of the $E$ polarization field in a dipole modulation model with $w_{1}=2.5$.  The power is larger in the southern 
hemisphere.} 
\label{w1high}
\end{center}
\end{figure}

Once $w_1$ is specified, we can use our simulation pipeline to make realizations of
the polarization field.   An extreme example with $w_1=2.5$ is shown in Fig.~\ref{w1high}.
Note that the hemispherical power asymmetry carries through to the polarization. 
The WMAP temperature field however suggests a much weaker dipole modulation that
can only be detected statistically in the polarization.
We will base our parameter choices on the analysis of Ref. \cite{eriksen_banday}
which found  $w_1 = 0.23 \pm 0.07$ with $w_1=0$ excluded
at the 99\% CL.
For illustrative purposes we will take a reionization model with 
$\tau = 0.09$ where hydrogen is fully ionized out to a maximum redshift and neutral
thereafter. We test sensitivity  to the reionization history
in \S \ref{sec: caveats}.

\subsection{Dipole Modulation Statistics} \label{sec: estimator_w1}

The dipolar modulation in the potential field gives rise to couplings between $\ell$ and $\ell^{\prime} = \ell \pm 1$ multipole moments in not only $TT$
\begin{equation}
\langle T_{\ell m}^* T_{\ell+1,m} \rangle \propto w_1 \,,
\end{equation}
but also in all of the two point combinations involving $E$
\begin{equation}
\langle X_{\ell m}^*Y_{\ell+1,m} \rangle \propto w_1\,,
\end{equation}
specifically $XY \in \{ TE, ET, EE \}$.  We derive the proportionality coefficients in 
Appendix \ref{sec: Unbiased estimator}.    Each multipole pair then forms a noisy
estimator of the dipole modulation parameter $w_1$.   Even in the absence of
detector noise and foregrounds, the cosmic variance of the temperature and
polarization fields provide noise to the measurement.
Eqs.~(\ref{AlmXY})
and (\ref{w1XY}) give the unbiased minimum variance combinations of the multipole
pairs as estimators of $w_1$.

\begin{figure}[htbp]
\begin{center}
\includegraphics[width=3.5in]{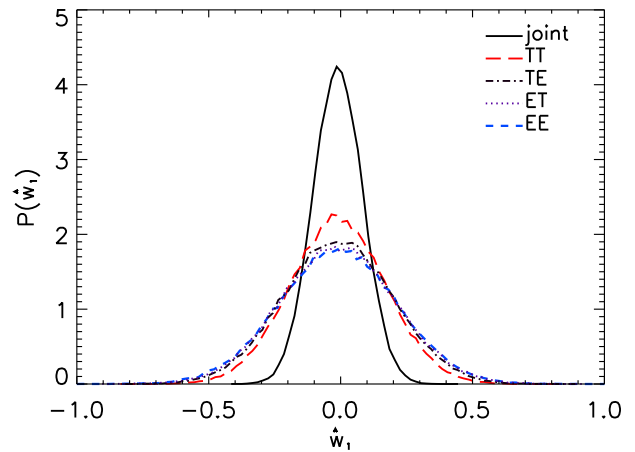}
\caption{Probability distribution of the dipole modulation estimators $\hat{w}_{1}^{XY}$ for 
a model with $w_{1}=0$
where $XY$ represents the two fields whose $\ell$ and $\ell+1$ multipoles are correlated: $TT$ (red long-dashed curve), $TE$ (black dotted-dashed curve), $ET$ (violet dotted curve) and $EE$ (blue thick dashed curve). The minimum variance joint estimator $\hat{w}_{1}$ is shown as the
solid curve.  
We have assumed cosmic variance limited measurements and $\ell_{\rm max}=11$ for
all estimators.} 
\label{w1_estimators}
\end{center}
\end{figure}

The first question we ask is: if the observed dipole asymmetry at
$\ell \sim 20-40$ is a statistical fluke and that the true ensemble average is
 $ w_1 =0$, how well can
temperature and polarization at $\ell \lesssim 10$ 
rule out a finite value of $w_1$ near the best fit of $w_1 \approx 0.2$?
Under this assumption, we show the distribution of the $XY$ estimators and the
joint estimator in Fig.~\ref{w1_estimators}.  These distributions were generated
from  $10^5$ 2D simulations of the unmodulated temperature and polarization fields
in accordance with the power spectra of Eq.~(\ref{unmodtemp}), (\ref{standard_polarization})
and (\ref{unmodte}).

Given $TT$ alone, the hypothesis that $w_1 \sim 0.2$ can only be weakly tested.
Specifically, a value of $w_{1}\ge 0.2$ can be tested
at the $86.4\%$ confidence level (CL) by $TT$.  
Each of the polarization estimators can in principle test dipole modulation at a comparable
level: $w_1 \ge 0.2$ can be tested at $81.6\%$ confidence by $EE$, at $83.4\%$ CL by $TE$ and at $82.1\%$ CL by $ET$.  The combination of the four estimators however is more powerful:
 $w_{1}\ge 0.2$ can be tested at $98.4\%$ CL by the joint estimator.  For reference, the
 central value of \cite{eriksen_banday}, $w_{1}=0.23$, yields a 99.3\% CL test.
 Furthermore the $95\%$ upper limit on $w_{1}$ achievable with the joint estimator is $w_{1}\le 0.15$.

\begin{figure}[htbp]
\begin{center}
\includegraphics[width=3.5in]{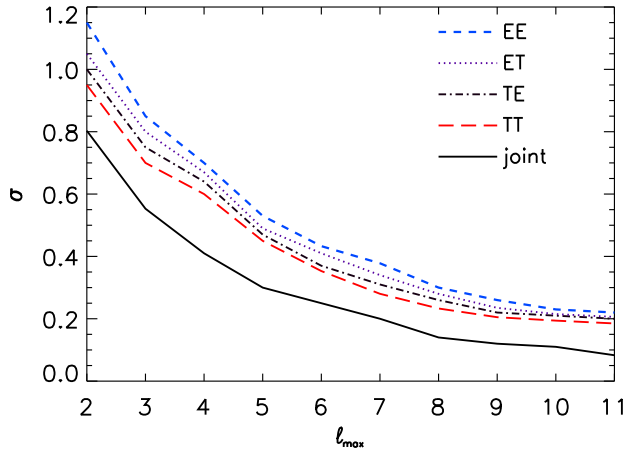}
\caption{Standard deviation of the individual estimators $\hat{w}_{1}^{XY}$ and the joint estimator 
(black) as a function of maximum multipole pair employed in the estimators $\ell_{{\rm max}}$.
The individual $XY$ pairs are $TT$ (red long-dashed curve), $TE$ (black dotted-dashed curve), $ET$ (violet dotted curve) and $EE$ (blue dashed curve) from the analytic calculation of the Appendix.}
\label{stddev_allcases}\end{center}
\end{figure}

Most of the polarization information comes from multipoles in the range of 
$\ell \sim 4-8$.   In Fig.~\ref{stddev_allcases} we show the rms fluctuation in the estimators as
a function of $\ell_{\rm max}$, the maximum multipole employed in the estimators.  Note
that the estimators involving polarization  saturate by $\ell \sim 10$.

\begin{figure}[htbp]
\begin{center}
\includegraphics[width=3.5in]{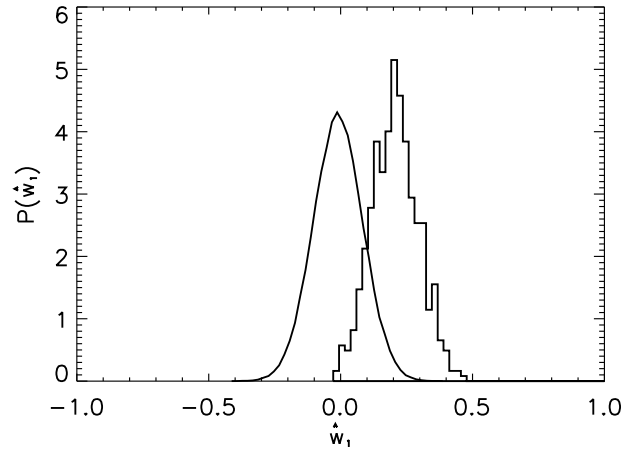}
\caption{Histogram of the joint estimator $\hat{w}_{1}$ coming from $545$ 3D realization of the
dipole modulation model with $w_{1}=0.2$ and  $\tau=0.09$. For comparison, the probability 
distribution for the isotropic $w_{1}=0$ model from Fig.~\ref{w1_estimators} is also plotted (smooth curve).}
\label{hist_3d}
\end{center}
\end{figure}

The next question we ask is: given a true model of $w_1 = 0.2$, how well can an isotropic model
of $w_1=0$ be excluded?    Fig.~\ref{hist_3d} shows the distribution of $w_{1}$ coming from $545$ realizations of a 3D modulated model (histogram, $\ell_{\rm max}=11$).  In these simulations $w_1 \le 0$ occurred
in less than $1\%$ of the realizations.  Furthermore  the mean of the samples $\langle \hat w_1 \rangle= 0.20$
is consistent with being unbiased and the standard deviation  $\sigma= 0.08$ is consistent
with the expectations from Fig.~\ref{stddev_allcases}. 

\subsection{Model Dependence and Noise}
\label{sec: caveats}

These tests are only weakly sensitive to the assumed reionization history for a cosmic
variance limited observation.   
The main difference is that a linear modulation produces a weaker effect for nearby 
sources of polarization and hence the signal decreases with $\tau$. 
As an example, we calculate the level of detectability for an unphysically small optical depth of $\tau=0.0279$. Even for this extreme case, values of $w_{1}\ge 0.2$ are ruled out at the $96.4\%$ CL  by the joint estimator. An intermediate case of $\tau=0.055$ rules out values of $w_{1}\ge 0.2$ at the $98\%$ CL.

Because most of the signal comes from low multipoles in the polarization $\ell \sim 4-8$,
the cosmic variance limit described in the previous section is within the statistical
reach of the Planck satellite.  
For definiteness we take the $143$ ${\rm GHz}$ and $217$ ${\rm GHz}$ frequency channels
and combine them into an effective white noise power spectrum of
$N_{\ell}^{EE} = (67.6  \mu$K-arcmin$)^2$ at low multipoles.   Noise in the temperature field is
negligible compared with cosmic variance.  
Quantitatively, for a reionization history with optical depth $\tau = 0.09$, and the noise level considered for Planck, values of $w_{1}\ge 0.2$ are ruled out at $97.8\%$ CL by the joint estimator $\hat{w}_{1}$.
A space-based experiment with no foregrounds, $5$ frequency channels and a noise level of $18\mu$K-arcmin
per frequency channel 
would essentially saturate the cosmic variance bound in each channel.
Given these results, foregrounds and systematics will probably be the limiting factor in how well
dipole modulation can be tested by polarization.   We leave this subject to a future work.

Finally, the model spectrum assumes that the temperature anisotropy comes purely from
the Sachs-Wolfe effect.  A $\Lambda$CDM model would predict that the ensemble 
averaged quadrupole has a substantial contribution from the so-called Integrated Sachs-Wolfe
(ISW) effect from the decay of potentials during the acceleration epoch.  Under 
$\Lambda$CDM, the measured
 low power
quadrupole anisotropy could indicate that either the ISW contributions to the quadrupole
are by chance small on our sky or that there is a chance cancellation between the two contributions.
Finally, it could indicate an alternate model of the dark energy where the ISW contributions
are smaller \cite{Caletal97,Hu98,BeaDor04} or designed to cancel the Sachs-Wolfe
effect \cite{GorHu04}.   In all of these explanations, the polarization field is unaffected.

Nonetheless, the dipole modulation tests involving the temperature field are slightly weakened
if the ISW contributions to $\Lambda$CDM are added back as a source of noise.  
Correspondingly the joint estimators also weaken.
For example, for a reionization model with optical depth $\tau = 0.09$ the joint estimator
can test values of $w_{1}\ge 0.2$ at the $97.6\%$ CL.

In summary, prospects for future polarization tests of dipole modulation at the $w_{1}\approx 0.2$ level at
the $\sim$98\% CL are fairly robust to model assumptions.

\section{Quadrupolar Modulation}
\label{sec: quadrupolar}

de Oliveira-Costa {\it et al.}~\cite{oliveira} pointed out that there are several striking features of the 
temperature dipole, quadrupole and octopole in the WMAP data.  The octopole is highly planar and 
defines a preferred axis that is near the dipole axis.  Moreover the quadrupole is also weakly
planar in the same coordinate system.  

 Gordon {\it et al.}~\cite{wayne} subsequently showed that  a quadrupolar modulation with
 a preferred axis that is by chance near the dipole direction can make these features substantially more likely.  
Specifically,  we will take a
modulation to the potential field of the form of Eq. (\ref{general_model1}), with
\begin{equation} \label{quad_h}
h({\bf x})=w_{2} Y_{20}(\frac{z}{D_{{\rm rec}}})\,.
\end{equation}
Note that the $z$-axis required here is {\it not} the same as in the dipole modulation
case.  Since $Y_{20}(z/D_{{\rm rec}})=(1/4)\sqrt{5/\pi}\left[ 3(z/D_{{\rm rec}})^{2} -1\right]$, 
this introduces a quadratic term to the spatial modulation.

These anomalies are confined to the low multipoles and so the modulation must also
be confined to low multipoles.  For the angular modulation considered in \cite{wayne},
this was simply imposed as a sharp suppression in multipole space of the $g_1$ field.  
Spatial modulation is more complicated in that even a sharp suppression in Fourier
space is broadened by projection effects.    Given these considerations, we
take a two field model defined   by
\begin{eqnarray} 
 \frac{k^{3} P_{g_1}(k)}{2 \pi^{2}} &=& {9 \over 25}  \delta_{\zeta}^{2} f^{2}, \qquad k_{1} < k < k_{2}\,,  \label{c11} \\
 \frac{k^{3} P_{g_2}(k)}{2 \pi^{2}} &=& {9 \over 25} \delta_{\zeta}^{2}, \hphantom{f^{2}}\qquad k > k_{3} \,, \label{c22}
\end{eqnarray}
where the normalization $\delta_{\zeta}$ is given as in the dipole modulation model by
the WMAP normalization.   We introduce a scaling parameter $f$ that controls the 
amount of power at low $k$ relative to this normalization. 
We begin our analysis
 by finding regions in the parameter space of $\{w_{2},k_{1},k_{2},k_{3},f\}$ that maximize the probability of
 the observed quadrupole-octopole alignment.

\subsection{Maximizing Temperature Alignments}\label{sec:alignments_temperature}

To quantify the quadrupole-octopole alignment, de Oliveira-Costa {\it et al.}~\cite{oliveira} 
introduced the normalized angular momentum
along a given axis $(\theta,\phi)$
\begin{equation}
\hat{L}^{2}_{\ell}(\theta,\phi) = \frac{\sum_{m}m^{2}|T_{\ell m}^
\prime |^{2}}{\ell^{2}\sum_{m}|T_{\ell m}^\prime|^{2}} \,,
\end{equation}
where the $T_{\ell m}^\prime$ are the multipole moments of the temperature field in that
preferred frame.  More specifically, from the moments defined in galactic coordinates $T_{\ell m}$,
the rotated ones are
\begin{equation}
T^{\prime}_{\ell m^{\prime}}=\sum_{m}T_{\ell m}D_{m m^{\prime}}^{\ell}(-\psi,-\theta,-\phi)\,,
\label{Tlmrot}
\end{equation}
where the Euler angles are $\psi \in \left[0,2\pi\right]$, $\theta \in\left[0,\pi\right]$, 
and $\phi \in\left[0,2\pi\right]$ and $D_{m m^{\prime}}^{\ell}(-\psi,-\theta,-\phi )$ is the Wigner rotation matrix
 \cite{copi,wayne}. Note that $\hat L^2_\ell$ is invariant under the final rotation by
$\psi$.

In particular, the statistic
\begin{equation}
\hat{L}^{2}_{23} = \frac{1}{2} (\hat{L}_{2}^{2} + \hat{L}_{3}^{2})\,,
\end{equation}
captures both the alignment between the quadrupole and octopole, and the planarity of the octopole.  The preferred axis is then chosen as the one which maximizes this statistic.
For the ILC map of the 3-year WMAP data corrected for the kinematic quadrupole,
 $\hat{L}^{2}_{23} = 0.943$.  This is slightly
smaller than for the 1-year WMAP data in the TOH map 
\cite{tegmark},  used in previous studies:
$\hat{L}^{2}_{23}  =0.96$ \cite{copi,wayne}.   For an unmodulated
model, $\hat{L}^2_{23} \ge 0.943$ occurred  in only $0.85\%$ of $10^4$ 2D realizations.

Quadrupole modulation can dramatically increase the probability of a high angular
momentum statistic.   To find the parameters that maximize this probability,
we first vary the parameters $\{w_{2}, k_{1}, k_{2} \}$ with $k_3 \rightarrow \infty$ and find the set that gives maximum number of realizations with $\hat{L}^{2}_{23} > 0.943$.
Taking $k_3 \rightarrow \infty$ eliminates the contributions from the unmodulated
$g_2$ field that could otherwise destroy alignments at low multipoles.
The alignment probability then is also independent of $f$.

We then generate $10^4$ 2D Monte Carlo (MC) realizations for  each of the modulated models. For each realization, the estimator $\hat{L}^{2}_{23}$ is computed on a grid of $10^{8}$ positions on the celestial sphere by performing the rotation of $T_{\ell m}$ in 
Eq.~(\ref{Tlmrot}). Taking the maximum value of $\hat{L}^{2}_{23}$ attained for each realization, we compute for each model the percentage of realizations with $\hat{L}_{23}^{2}>0.943$.

The region of parameter space that corresponds to the highest alignments 
is centered around $w_2 \approx -7$, in agreement with \cite{wayne}.
Furthermore, wavenumbers that best project onto a multipole $\ell$ correspond to
$k D_{\rm rec} \sim \ell$; more specifically
$k D_{\rm rec} \approx 4.2$ for the region  that best projects onto the quadrupole and octopole at 
the distance to recombination.
Furthermore keeping $k_1$ and $k_2$ in a narrow
range around this value eliminates contamination from the modulation
of longer and shorter wavelengths in the quadrupole and octopole. 

For definiteness, we take  $w_{2}=-7$, $k_{1}D_{\rm {rec}}=4.2$ and $k_{2}D_{\rm{rec}}=4.23$
though any $(k_2-k_1) D_{\rm rec} \lesssim 0.05$
would produce similar results due to
unavoidable projection effects.    With this choice, 
$21.8 \%$ of realizations have $\hat{L}^{2}_{23} > 0.943$.  This number is smaller than that
quoted in \cite{wayne} due to these projection effects but is still a factor of $25.6$ larger than
the probability in unmodulated models.

While taking $k_3 \rightarrow \infty$ maximizes alignments, it does not provide for a realistic
model.   The absence of high wavenumbers would be in conflict with the observed fluctuations
at higher multipole moments.  
This is exhibited in a poor ILC likelihood 
\begin{equation}
\mathcal{L}({\bf T}|{\bf C}) = {1\over(2 \pi)^{N/2}\sqrt{\rm det {\bf C}}}{\rm exp}[-{1 \over 2} {\bf T}^{\dagger}{\bf C}^{-1}{\bf T}]\,, \end{equation}
where ${\bf T}$ is the $N$ element vector of observed $T_{\ell m}$ values and ${\bf C}$ is their signal plus
noise covariance matrix.    To remedy this problem we keep the first field fixed to the parameters above but add in the second field at a finite
$k_3$ and maximize the likelihood over their relative amplitude $f$.

\begin{figure}[htbp]
\begin{center}
\includegraphics[width=3.5in]{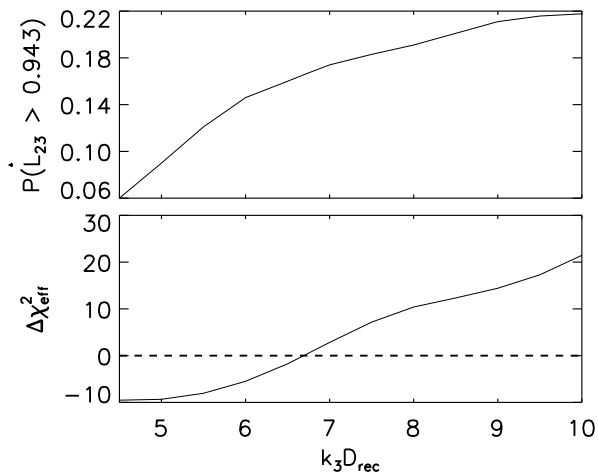}
\caption{Effect of the unmodulated field on alignments and goodness of fit of 
quadrupole modulation models.
The upper panel shows the fraction of realizations that have a higher angular momentum statistic $\hat{L}_{23}^2$ than WMAP 3-year ILC as a function of $k_{3}D_{\rm rec}$
which controls the contribution of the unmodulated field to the low multipoles.  
The dependence of $\Delta \chi^2_{{\rm eff}}$ 
(relative to $\Lambda$CDM, dashed line) is shown in the lower panel.  }
\label{chi2_align}
\end{center}
\end{figure}

\begin{figure}[htbp]
\begin{center}
\includegraphics[width=3.5in]{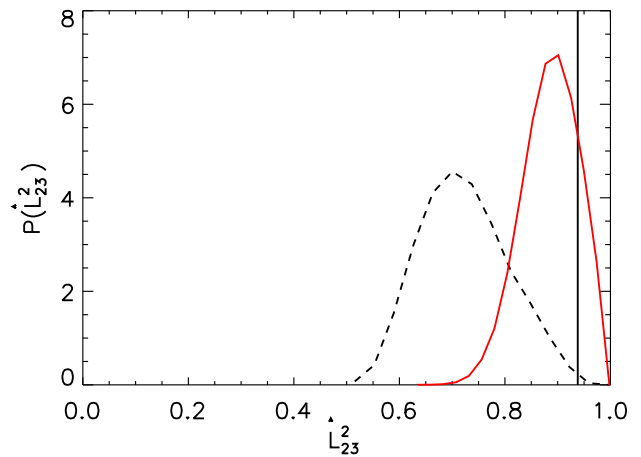}
\caption{Distribution of $\hat{L}^{2}_{23}$ (red solid line) for the
quadrupolar modulation model of \quadmod. 
 $16.6 \%$ of the $10^{4}$ realizations
 of this model have $\hat{L}^{2}_{23} > 0.943$.
The distribution for the isotropic case is shown as a dashed line for comparison, where 
$0.85\%$ have  $\hat{L}^{2}_{23}$ larger than the WMAP ILC value: 0.943 (vertical line).} \label{L23T}
\end{center}
\end{figure}

Fig.~\ref{chi2_align} (lower panel) shows the dependence of $\chi^2_{\rm eff} \equiv -2\ln \mathcal{L}$ on $k_3 D_{\rm rec}$ for $\ell=2-5$ relative to $\Lambda$CDM (see below Eq.~(\ref{LCDMmodel})).
Decreasing this parameter creates a better fit to the data at the expense of decreasing the
probability of the quadrupole-octopole alignment (upper panel).  As a compromise we choose
a model with $\Delta\chi^{2}_{\rm eff}=0$ that neither improves nor worsens the fit.
The model parameters then become
\begin{align}\label{set_max}
 \quad k_{1}D_{\rm {rec}}&=4.2\,, \quad k_{2}D_{\rm{rec}}=4.23\,, \quad
k_{3}D_{\rm rec}=6.697\,,\nonumber\\
 \quad w_{2}&=-7\,, \quad f=5.77\,, \quad \tau=0.09 \,,
\end{align}
for which the probability of $\hat{L}^{2}_{23} > 0.943$ is $16.6\%$ (see Fig.~\ref{L23T}).

\begin{figure}[htbp]
\begin{center}
\includegraphics[width=3.5in]{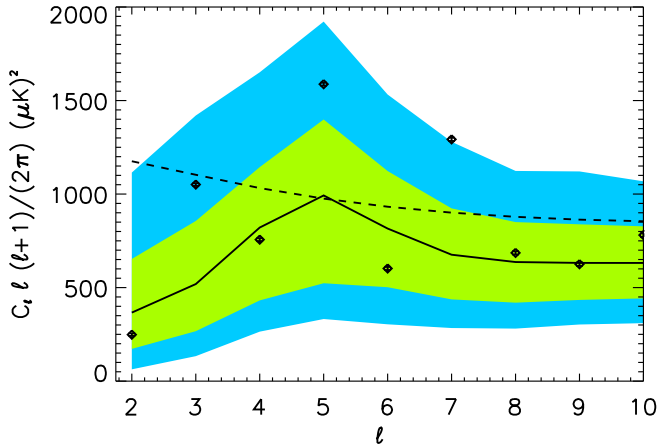}
\caption{Temperature angular power spectrum of the quadrupolar modulation
 model 
of
\quadmod\
 (solid line). The WMAP ILC data with noise errors are also shown. The solid bands correspond to the $68\%$ and $95\%$ cosmic variance confidence regions of the model. 
 The dashed line represents the $\Lambda$CDM model against which this model is compared
 in Fig.~\ref{chi2_align}.} \label{Cl_total}
\end{center}
\end{figure}

Fig.~\ref{Cl_total} shows the temperature angular power spectrum of this model
along with the WMAP 3-year temperature data for the first $10$ multipoles. The bands shown here represent the $68\%$ and
$95\%$ cosmic variance confidence regions for this model taken from 
$10^4$ 2D realizations of the temperature field.   Note that the cosmic variance is larger for the
modulated model due to the $m$-dependence of the power.

\begin{figure}[htbp]
\begin{center}
\includegraphics[width=4.in]{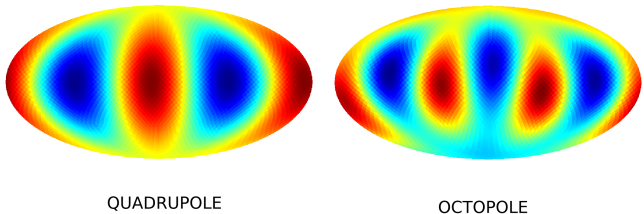}
\caption{Quadrupole and octopole of the temperature field from one realization of the quadrupole modulation model of \quadmod.   The normalized angular momentum here is $\hat{L}_{23}^{2}=0.94$.  The range goes from  $-50\mu$K to $50\mu$K for both cases.}\label{align_qo}
\end{center}
\end{figure}

In Fig.~\ref{align_qo} we show a realization of the model that has a high normalized angular momentum $\hat{L}_{23}^{2}=0.94$, shown in the frame of the modulation, illustrating how the modulation takes a random isotropic sky and aligns the quadrupole and the octopole.

\begin{figure}[htbp]
\begin{center}
\includegraphics[width=3.5in]{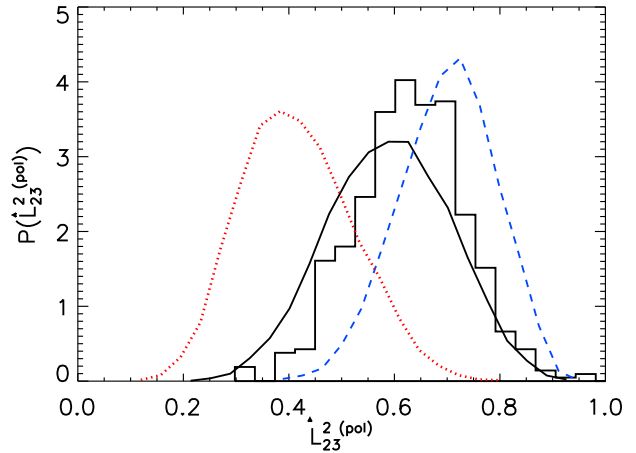}
\caption{Histogram of $\hat{L}_{23}^{2 (\rm{pol})}$ for $571$ 3D realizations   of 
the quadrupole modulation model of \quadmod\ that also have
$\hat{L}_{23}^{2} \ge 0.943$.   For comparison we also show the distribution for
isotropic random polarization fields (red dotted lines).  
Isotropic realizations constrained to the $T_{\ell m}$ values of the ILC map have 
a distribution shifted to  higher alignment values given the temperature-polarization
correlation with just the Sachs-Wolfe effect (blue dashed lines) and with $\Lambda$CDM
ISW contributions added  (black solid curve).} \label{L23pol_const}
\end{center}
\end{figure}

\subsection{Polarization Angular Momentum} \label{sec: Alignments_polarization}

With the quadrupolar modulation model set by \quadmod\ to maximize temperature alignments and fit
the WMAP temperature data, we can
now use our simulations to make polarization predictions.  

We first examine the normalized angular momentum statistic
of the polarization field.   We define this statistic in the same way as for the temperature field
\begin{equation}
\hat{L}_{l}^{2 (\rm{pol})}=\frac{\sum_{m} m^{2}|E_{\ell m}|^{2}}{\ell^{2}\sum_{m} |E_{\ell m}|^{2}} \,,
\end{equation}
with the combined quadrupole and octopole statistic being
\begin{equation} \label{l23polariz}
\hat{L}_{23}^{2 (\rm{pol})} = \frac{1}{2}(\hat{L}_{2}^{2(\rm{pol})} + \hat{L}_{3}^{2(\rm{pol})})\,.
\end{equation}
There is one important difference between this statistic and the temperature based one.
In the case of polarization predictions, the temperature field has already defined the preferred
axis against which we should measure the polarization angular momentum.  Consequently,
no search over directions is required.

\begin{figure}[htbp]
\begin{center}
\includegraphics[width=3.5in]{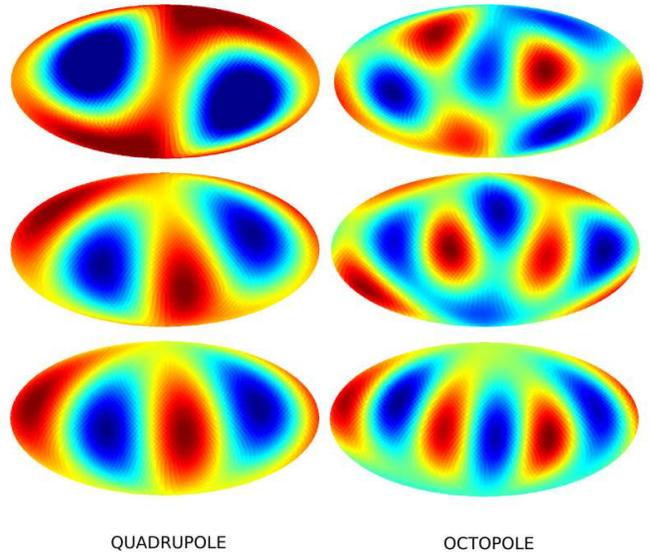}
\caption{Realizations of the quadrupole (left column) and octopole (right column) of the $E$ component of the polarization field in order of increasing alignment:  isotropic case with $\hat{L}_{23}^{2 (\rm{pol})}=0.46$ (first row), quadrupolar modulation model with parameters given by \quadmod\
with $\hat{L}_{23}^{2 (\rm{pol})}=0.67$ (second row) and an isotropic Sachs-Wolfe model constrained to match the WMAP ILC temperature multipoles with $\hat{L}_{23}^{2 (\rm{pol})}=0.9$ (third row). The color scale goes from $-0.3\mu$K to $0.3\mu$K 
in all panels.}
\label{Maps_quad_oct_E} 
\end{center}
\end{figure}

Fig.~\ref{L23pol_const} shows a histogram of
$\hat{L}_{23}^{2 (\rm{pol})}$ for
$571$ out of 3482 3D realizations of the modulated
model that have as high or higher temperature alignment than WMAP
 $\hat{L}_{23}^{2} > 0.943$.   We first compare this with the same statistic 
 for $10^5$ 2D isotropic realizations of an unmodulated polarization field
 (red dotted line).   
There is a tendency for the polarization field
 to be aligned compared with random isotropic realizations.
 An example of a typical realization is shown in  Fig.~\ref{Maps_quad_oct_E} (second row)
  and is to be compared with a typical isotropic realization (first row).
 Note the similarity between temperature alignments and $E$ polarization aligments
 in Figs.~\ref{align_qo} and~\ref{Maps_quad_oct_E}.  Both have a deficit of power near the poles
 in the frame of the modulation. However the polarization amplitude 
 $(Q^{2}+U^{2})^{1/2}$ is in fact the largest at the poles.  In this direction, the electron
 scattering sees the planar quadrupole in the transverse plane and scatters it into a maximal
 polarization.   
 Foregrounds which have high polarization where there is high intensity will not have
 this pattern.

 However,  the expected
 correlation between the temperature and polarization
 fields implies that there should be {\it some} tendency for polarization alignment in
  {\it any} explanation in which the alignment of the temperature field
  reflects contributions from the recombination surface, and not a foreground, secondary
   or systematic effect.  This includes an isotropic
 unmodulated model in which the temperature alignment is a chance occurrence.
 
 To quantify this tendency, we consider two isotropic models which have been conditioned to have
 $\hat{L}_{23}^2 = 0.943$ for the temperature field.  More specifically, we draw a polarization
 field that is consistent with the observed temperature field $T_{\ell m}$ and model 
 power spectra,
 \begin{equation}
E_{\ell m}= {C_{\ell}^{TE} \over C_{\ell}^{TT}}T_{\ell m} + \sqrt{C_{\ell}^{EE} - {(C_{\ell}^{TE})^{2}\over C_{\ell}^{TT}}}n_{\ell m}\,,
\end{equation}
where $n_{\ell m}$ is a complex Gaussian variate of zero mean with
$\langle n_{\ell m}^* n_{\ell m} \rangle = 1$ and $n_{\ell m}^* = (-1)^m n_{\ell,-m}$.

 The first is the Sachs-Wolfe model where
 all of the temperature anisotropy arises from recombination.  Under this assumption, the
 temperature alignments are a real feature of the recombination surface and the ISW effect in
 our local universe happens to be small in the quadrupole and octopole.  Fig.~\ref{L23pol_const} 
 (blue dashed lines) shows the polarization statistic for $10^5$ 2D constrained 
 realizations of this model.  
 An example of a highly aligned realization is shown in Fig.~\ref{Maps_quad_oct_E} (third row).

Compared with the modulated model, there is an even {\it higher} probability of
 polarization alignment here.  The reason is that in the former, the strength of the
 spatial modulation increases as the square of the
 distance.  The closer distance to reionization implies a substantially weaker modulation of the
 polarization field as compared with the temperature field.  Ironically, a very high value
 of $\hat{L}_{23}^{2 (\rm{pol})}$ would strongly disfavor the modulated model compared
 with the isotropic Sachs-Wolfe model.
The $95\%$ upper limit for the modulated model is $\hat{L}_{23}^{2 (\rm{pol})} = 0.77$.

 It is also possible that the alignment of the quadrupole and octopole does not represent an 
 aligned spatial
 configuration at all and is instead the result of a chance combination of the Sachs-Wolfe effect and ISW
 effects.  These effects arise from very different physical scales but are 
 seen in projection at the same angular scale.
 To illustrate this case, we draw $10^5$ 2D realizations of the $\Lambda$CDM model
 (see below Eq.~(\ref{LCDMmodel})) that
 are similarly conditioned to have $\hat{L}_{23}^2 = 0.943$ for the temperature field.  
 The histogram of the polarization statistic is shown in Fig.~\ref{L23pol_const} (black solid curve).
 The presence of a temperature contribution that does not correlate with polarization shifts
 the distribution downwards and decreases the probability of polarization alignments.
 The distribution is still distinct from the truly random isotropic simulations (red dotted lines).
The 95\% lower limit on this model is $\hat{L}_{23}^{2 (\rm{pol})} \ge 0.39$.

 The final possibility is  that the temperature quadrupole and temperature field is nearly
 all local secondary anisotropy 
 and that the Sachs-Wolfe effect is anomalously low (e.g. \cite{Vale:2005mt,Inoue:2006rd}).
 The polarization distribution would
 then look like the truly random isotropic simulations (see Fig.~\ref{L23pol_const},
 red dotted line).
In this case,  the polarization amplitude itself would also be
 anomalously low.
 
 In summary, a very low value of $\hat{L}_{23}^{2 (\rm{pol})} \lesssim 0.4$ with a normal
 polarization amplitude would suggest
 that the origin of the temperature alignments is not cosmological.  A very low value
 with suppressed polarization amplitude would imply a purely secondary 
 origin to the temperature
 alignment.  A very high value
 $\hat{L}_{23}^{2 (\rm{pol})} \gtrsim 0.8$ would rule out the modulated model and also
 imply that the ISW effect is nearly absent in the WMAP low multipoles.

 \begin{figure}[htbp]
\begin{center}
\includegraphics[width=3.5in]{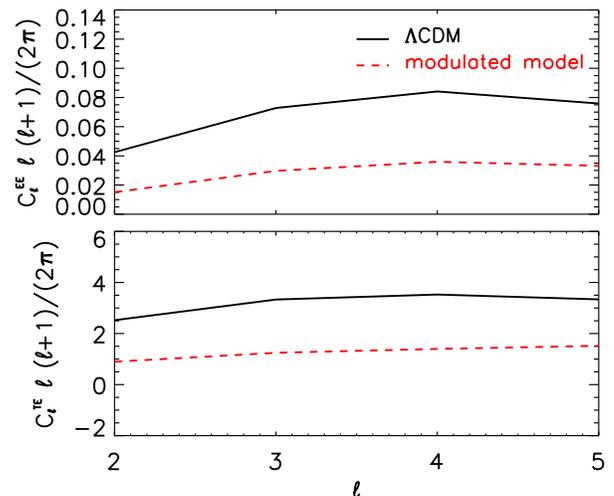}
\caption{$EE$ and $TE$ power spectra of the modulated model of \quadmod\  (red dashed lines) along with the $\Lambda$CDM power spectrum (black solid lines). The modulated model has a lower polarization at the same $\tau(=0.09)$. The units are  $(\mu$K$)^{2}$. }
\label{fiducial_modulated}
\end{center}
\end{figure}

 \subsection{Quadrupole Modulation Statistics} 

Beyond the  angular momentum of the quadrupole and octopole, a quadrupolar modulation
model
 makes further testable predictions.  These predictions however are more sensitive to the choice
of model parameters.

First, the angular power spectrum of the polarization and temperature cross correlation differs
in its relationship to the temperature power spectrum due to the modulation.  
In Fig.~\ref{fiducial_modulated}, 
we show these spectra for the model of  \quadmod.   Note that for the same optical depth $\tau$, the polarization 
spectra have less power.   This is a consequence of the decrease in modulation amplitude
at the closer distance to reionization.  Unfortunately, this signature is largely degenerate
with a change in the optical depth and so is not a unique signature of modulation.

Similar to dipole modulation, quadrupole modulation
generates a more unique signature in the 
correlation between neighboring multipole moments
\begin{equation}
\langle X_{\ell m}^*Y_{\ell+2,m} \rangle \ne 0 \,,
\end{equation}
where $XY \in \{ TT, TE, ET, EE \}$.  In our two field model, most of the modulation is
confined to the low multipoles and so the signal is mainly in $\ell +2 =4,5$.  
Note that an isotropic model that by chance has quadrupole-octopole temperature
alignments will still predict a zero expectation value for these quantities. 

Let us define a correlation statistic
\begin{equation}
 \hat s_{\ell m}^{XY}  = { {{\rm Re}( X_{\ell m}^*Y_{\ell+2,m} )} \over | X_{\ell m}||Y_{\ell+2,m}|} \,,
\end{equation}
which tests the phase alignment of the moments and takes on values from $-1 \le \hat s_{\ell m}^{XY} \le
1$.  In the absence of modulation, these statistics have zero mean but any given pair
has large cosmic variance.  Given a modulation model that
predicts
\begin{equation}
 s_{\ell m,{\rm mod}}^{XY} = \langle  \hat s_{\ell m}^{XY}  \rangle_{\rm mod} \,,
\end{equation}
we can construct a weighted sum 
\begin{equation}\label{stot_def}
\hat s ={ \sum_{\ell m, XY}  s_{\ell m,{\rm mod}}^{XY}
 \hat s_{\ell m}^{XY}  \over
  \sum_{\ell m, XY} \left( s_{\ell m,{\rm mod}}^{XY}\right)^{2}}\,,
\end{equation}
where $\ell=2,3$.
This statistic has the property that its expectation value is zero for
an isotropic  field and $1$ for the chosen modulation model.

\begin{figure}[htbp]
\begin{center}
\includegraphics[width=3.5in]{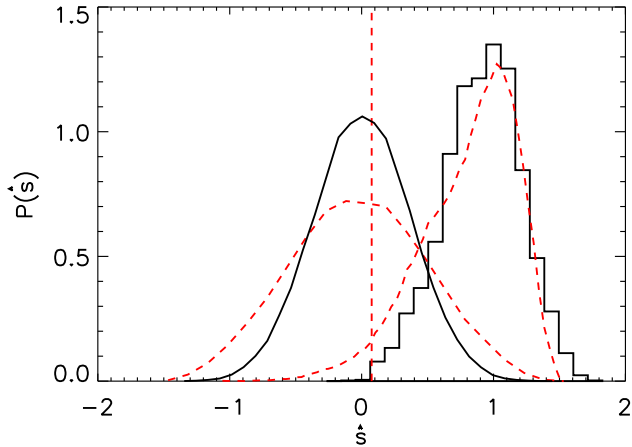}
\caption{Distribution of  $\ell \pm 2$ correlation statistic $s$.  Distributions
 on the right with $\langle \hat s \rangle =1$ represent the  modulated model of \quadmod\
 for the joint   estimator
 (black solid lines) and the temperature estimator (red dashed lines).   Curves on the left
 with $\langle \hat s \rangle =0$ represent the isotropic Sachs-Wolfe model in the same
 manner. The histogram comes from $3482$ 3D realizations of the modulated model
 and other distributions from $10^{5}$ 2D realizations. The vertical red dashed line represents the observed value for temperature only for this modulated model, using the WMAP ILC map: $\hat s = 0.076$. }
\label{model1}
\end{center}
\end{figure}

We begin by considering the $XY=TT$ case and the modulated model of 
\quadmod\ as in the previous section.  
The distribution of $\hat s$ is shown in Fig.~\ref{model1}
(dashed lines, right curve) from $10^5$ 2D simulations.  
The WMAP ILC data has $\hat s=0.076$  and disfavors the modulated model but not at high significance: $3.5\%$
of the modulated simulations have $\hat s$ less than the data. Conversely the data would be a typical
realization of an isotropic model (see Fig.~\ref{model1}, dashed lines, left curve).
Likewise an $\hat s \le 0$ occurred in $2.5\%$ of the modulated simulations,
and an $\hat s \ge 1$  occurred in  $2.3\%$ of the isotropic simulations.
These two tails of the respective distributions represent the power of the statistic
to distinguish typical realizations of the models.

The addition of polarization information would make these tests definitive by
reducing the probability in the tails (see Fig.~\ref{model1}, solid curves).   When all fields are 
considered $XY \in \{TT,EE,TE,ET\}$, the probabilities in the tails drop to 
 $\hat s \le 0$ in only  $0.09\%$ of the $3482$ modulated 3D simulations
and $\hat s\ge 1$  in $0.24\%$ of the $10^5$ 2D isotropic simulations.
The latter probability assumes a Sachs-Wolfe only model for the temperature field.
If the ISW contribution of $\Lambda$CDM is factored in, the probability increases slightly
to $0.34\%$.

Consequently a quadrupolar modulation with parameters given by \quadmod\ will be 
definitively tested through the $s$ statistic by polarization measurements that approach the cosmic variance limit
for $\ell \le 5$.  However a model with a lesser amount of modulation at 
$\ell =4,5$ may evade such constraints.  The amount of modulation in these
multipoles is controlled by $k_3$ and hence $\hat s$ 
mainly constrains this parameter.  In the model
we have been considering, $k_3 D_{\rm rec}=6.697$.  Lowering this value increases
the contribution of the unmodulated field for $\ell \le 5$. But the price is to make 
the quadrupole-octopole alignment less likely (see Fig.~\ref{chi2_align}).  

As a specific example,
for
$k_{3}D_{\rm rec}=5.16$ and $f=5.13$, the probability of $L_{23}^2>0.943$ decreases from
16.6\% to 10\% (see Fig.~\ref{chi2_align}).  On the other hand, it improves the fit to the WMAP data by making 
$\Delta\chi^2_{\rm eff}=-9$ relative to $\Lambda$CDM.
Correspondingly, the $s$-statistic also weakens in its discriminating power.
For the $TT$ case, the WMAP ILC data has $\hat s=0.065$. For this case, $10.6\%$ of $2521$ simulations have a smaller value of $\hat{s}$. Similarly, a value of $\hat s \le 0$ occurred in $9.1\%$ of the modulated simulations, and $\hat s \ge 1$ occurred in $10.6 \%$ of the isotropic simulations.
With the addition of polarization information these tests become more significant:
 $\hat s \le 0$ only occurred in only $0.24\%$ of the modulated simulations
and  $\hat s\ge 1$  only occurred in in $0.50\%$ of the isotropic simulations.

In summary, while the $s$-statistic cannot rule out  quadrupolar modulation in general,
it can severely limit the amount of modulation allowable at $\ell \le 5$ and hence
test its ability to explain the quadrupole-octopole alignment.

\section{Discussion} \label{sec: discussion}

Statistical anomalies in the large-scale CMB temperature sky measured by WMAP have been suggested as possible evidence for the violation of statistical isotropy on large scales.  In any physical model for such a violation, there are testable consequences for
the CMB polarization field.   

We have developed a radiative-transfer simulation-based approach
to predicting these consequences.  We have applied this approach to modulation models
of broken isotropy.  In this class of models, a long-wavelength modulation of shorter-wavelength
fluctuations breaks isotropy locally, while the universe remains globally isotropic and homogeneous.

A linear or dipolar modulation can be used to explain the observed power asymmetry between
the northern and southern ecliptic hemispheres at multipole moments $\ell \sim 20-40$ \cite{eriksen_hansen,spergel,gordon, eriksen_banday}. 
Linear modulation produces a dipole coupling that correlates 
multipole moments $\ell$ to $\ell \pm 1$ in all of the two point functions of the temperature
and $E$ polarization field.   Using an analytic framework developed in the Appendix, we
derive an estimator of the modulation amplitude from these fields from these
correlations.   

We find that
the addition of polarization information substantially improves the ability to test
for dipole modulation at the lowest multipoles $\ell \lesssim 10$.  
Under a wide range of assumptions, the dipole modulation explanation of the
hemisphere asymmetry can be tested  at the 98\% CL or greater.
These tests will require near cosmic variance limited polarization measurements
at the low multipoles.   Such measurements are within the statistical reach of the {\it Planck}
satellite but systematic contamination from foregrounds will have to be sufficiently controlled.

A quadratic or quadrupolar modulation in the temperature field can  be used to 
explain the alignment of the quadrupole and octopole \cite{oliveira,wayne}.
We generalize the considerations of Ref.~\cite{wayne} and develop full 3D models
that enhance the probability of alignments by a factor of $\sim 20$.  

These models predict that the quadrupole and octopole of the $E$-polarization 
will tend to align as well.  However alignment of the polarization mainly tests that
the temperature alignment is cosmological and originates from recombination,
not the specifics of the
modulation model.  As such its main use will be to reject explanations of 
the alignments due to foregrounds or purely local secondary anisotropy.
In fact a very high alignment, comparable to the temperature field, is not 
expected in the modulation model due to the smaller strength of the modulation 
at the distance to reionization.   If such an alignment is seen then it
is more likely that it arises from a chance configuration of the temperature field
at recombination.

Quadrupolar modulation models can be distinguished in a manner similar
to the dipole modulation but through a correlation between multipoles $\ell$ and
$\ell \pm 2$ in the temperature and polarization fields.  We find that the combination
of temperature and polarization information out to $\ell \le 5$ can strongly constrain
the level of modulation in the low multipoles.  As a consequence, polarization measurements
can in principle rule out models  that make the temperature
alignments more than a few percent likely $\sim 99\%$ CL or stronger.

In general, matching signatures of broken isotropy in the polarization provide
a powerful test of competing explanations of the temperature anomalies.  
Explanations that involve foregrounds (e.g. \cite{Slosar:2004xj}) or local gravitational effects 
\cite{Inoue:2006rd,Vale:2005mt} would tend to predict different
signatures.   This is especially true since matching anomalies in the $E$ polarization 
are non-local in the Stokes parameters. Possible experimental systematics would be specific to a given experiment, and would thus have to be analyzed in a case by case basis. Therefore we do not consider them in this paper.

The examples we have studied illustrate that polarization in the future will provide a fruitful testing
ground for explanations of the observed temperature anomalies.  As predictions of 
anomaly models, polarization statistics move beyond the {\it a posteriori} inferences
that currently dominate the field.

\acknowledgments 

We thank Alexander Belikov, Colin Bischoff, Olivier Dor$\Acute{\rm e}$, 
Christopher Gordon, Gary Hinshaw,   Dragan Huterer, Michael Mortonson,
 Kendrick Smith, and Glenn Starkman
for useful discussions.
We acknowledge use of the HEALPIX package and the {\it Legacy Archive for Microwave Background Data Analysis} (LAMBDA). 
This work was supported by the KICP through the grant NSF PHY-0114422.
HVP was additionally supported by NASA through Hubble Fellowship grant \#HF-01177.01-A awarded by the Space Telescope Science Institute, which is operated by the Association of Universities for Research in Astronomy, Inc., for NASA, under contract NAS 5-26555. 
WH was additionally supported by the DOE  through 
contract DE-FG02-90ER-40560 and the David and
Lucile Packard Foundation.

\appendix

\section{Analytic Modulation Framework}

In this Appendix, we construct an analytic framework for modulated temperature
(\S \ref{sec: modulated_temp_statistics})
and polarization statistics (\S \ref{sec :general_polarization})
and evaluate them explicitly for dipole modulation.   We use these analytic calculations
to cross check numerical results in \S \ref{sec: Codetests} and to construct
estimators of the dipole modulation in 
\S  \ref{sec: Unbiased estimator}.

\subsection{Modulated Temperature Field} \label{sec: modulated_temp_statistics}

For the temperature field, 
the effect of modulation can be viewed as either the direct projection of the
modulated (convolved in Fourier space) field or the convolution in harmonic
space of the multipole moments of the original fields.
The latter perspective is taken in Ref. \cite{wayne} for the temperature field.  We show in this section that the former is equivalent and more closely resembles the derivation
required for polarization in \S \ref{sec :general_polarization}. 

For notational simplicity, we will derive the contribution to the spectrum due only to the modulated piece of the model of Eq.~(\ref{general_model1}) so that
\begin{equation} \label{general_model}
\Phi ({\bf x})=g({\bf x})\left[1+h({\bf x})\right] \,.
\end{equation}
The unmodulated piece can be added to these results without loss of generality.

The Fourier moments of $\Phi$ are given by a convolution of the two fields $g$ and $h$,
\begin{eqnarray} \label{convolutionapp}
\Phi({\bf k}) &=& g({\bf k})  + \int \frac{d^{3}k^{\prime}}{(2 \pi)^{3}}g({\bf k}^{\prime}) h({\bf k}-{\bf k}^{\prime}) \,,
\end{eqnarray}
and the two point function of the field $\Phi$ becomes
\begin{align}
\label{phi2ptapp}
\langle\Phi^{*}({\bf k})\Phi({\bf k}^\prime)\rangle
= &  (2\pi)^3 \delta({\bf k}-{\bf k^{\prime}}) P_{g}(k) \nonumber\\
& + \left[ P_{g}(k) +P_{g}(k^{\prime})\right]  h({\bf k}^{\prime}-{\bf k})\\
& +  \int {d^{3} \tilde k \over (2\pi)^{3}} P_{g}(\tilde k) 
h^{*}({\bf k}-\tilde{\bf k}) h({\bf k}^{\prime} -\tilde{\bf k})\,. \nonumber
\end{align}

Under the Sachs-Wolfe approximation of Eq.~(\ref{sachs-wolfe}),  the
spherical harmonic decomposition of the temperature field yields
\begin{eqnarray}
T_{\ell m} =- {1 \over 3} \int {d^{3} k \over (2\pi)^{3}} \Phi({\bf k}) 4\pi i^{\ell} j_{\ell}(k D_{\rm rec})
Y_{\ell m}^{*}(\hat {\bf k})\,,
\end{eqnarray}
which implies a two-point function of the temperature harmonics of
\begin{eqnarray}
\label{tlmtlmp}
\langle T_{\ell m}^{*} T_{\ell^{\prime} m^{\prime}}\rangle
 &=& {1 \over 9} \int {d^{3} k \over (2\pi)^{3}}
 \int {d^{3} k^{\prime} \over (2\pi)^{3}}
 \langle \Phi^{*}({\bf k}) \Phi({\bf k}^{\prime}) \rangle \nonumber\\
&& \times (4\pi)^{2} i^{\ell^{\prime}-\ell} j_{\ell}(k D_{\rm rec}) j_{\ell^{\prime}}(k^{\prime} D_{\rm rec}) \nonumber\\
&& \times Y_{\ell m}^{*}(\hat {\bf k}) Y_{\ell^{\prime} m^{\prime}}(\hat {\bf k}^{\prime}) \,.
\end{eqnarray}
Eqs. (\ref{phi2ptapp}) and  (\ref{tlmtlmp}) provide the analytic framework for
modulated temperature two-point statistics.

To make these considerations concrete,  let us explicitly evaluate Eq.~(\ref{tlmtlmp}) for 
dipole modulation.  
Dipole modulation can be taken
 to be a local approximation to a superhorizon scale Fourier mode,
\begin{eqnarray}\label{dipole_modulation}
h({\bf x}) &=& w_{1} \sqrt{3 \over 4\pi} {1 \over k_{0}D_{\rm rec}} \sin{({\bf k}_{0}\cdot {\bf x})} \nonumber\\
&\approx&  w_{1} \sqrt{3 \over 4\pi}{z \over D_{\rm rec}} \,,
\end{eqnarray}
where we choose $\hat {\bf k}_{0} \parallel {\bf z}$.
The Fourier representation of the field becomes
\begin{eqnarray}\label{equation_hfield}
h ({\bf k}) = {w_{1} \over 2 i}\sqrt{3 \over 4\pi}  {(2\pi)^{3}  \over k_{0}D_{\rm rec}}
[\delta({\bf k}-{\bf k}_{0}) - \delta({\bf k} + {\bf k}_{0})]\,.
\end{eqnarray}

Now let us substitute this expression into Eq.~(\ref{phi2ptapp}).  
The first term 
 is zeroth order in $h$ and is just the standard angular power spectrum
\begin{eqnarray}
\langle T_{\ell m}^{*} T_{\ell^{\prime} m^{\prime}}\rangle^{(0)} = 
 \delta_{\ell\ell^{\prime}}\delta_{mm^{\prime}}C_{\ell}^{TT}\,,
 \end{eqnarray}
 where
 \begin{eqnarray}
 \label{unmodtemp}
C_{\ell}^{TT} = {1 \over 9} \int {d k \over k} {k^{3}P_{g}(k)\over 2\pi^{2}} (4\pi) j_{\ell}^{2} (k D_{\rm rec})\,.\nonumber
\end{eqnarray}
The second term is first order in $h$ and is given by
\begin{eqnarray}
\langle T_{\ell m}^{*} T_{\ell^{\prime} m^{\prime}}\rangle^{(1)}
& = & {1 \over 9} \int {d^{3} k \over (2\pi)^{3}}
 \int {d^{3} k^{\prime}}
\left[ P_{g}(k) +P_{g}(k^{\prime})\right] \nonumber\\
&& \times {w_{1} \over 2 i}\sqrt{3 \over 4\pi}{1 \over k_{0}D_{\rm rec}}Y_{\ell m}^{*}(\hat {\bf k}) Y_{\ell^{\prime} m^{\prime}}(\hat {\bf k}^{\prime}) \nonumber\\
&& \times [\delta({\bf k}^{\prime}-{\bf k}-{\bf k}_{0}) - \delta({\bf k}^{\prime}-{\bf k} + {\bf k}_{0})]
\nonumber\\
&& \times (4\pi)^{2} i^{\ell^{\prime}-\ell} j_{\ell}(k D_{\rm rec}) j_{\ell^{\prime}}(k^{\prime} D_{\rm rec})\,. \nonumber
\end{eqnarray}

The simplification of these expressions requires $ j_{\ell^{\prime}}(k^{\prime} D_{\rm rec}) Y_{\ell^{\prime} m^{\prime}}(\hat {\bf k}^{\prime})$ to be expressed in terms of ${\bf k}$ where ${\bf k^{\prime}} = {\bf k} \pm {\bf k}_{0}$.
Assuming that $k_{0}/k \ll 1$, we can expand,
\begin{align} \label{taylor_temperature}
&{1 \over k_{0}D } \sqrt{3 \over 4\pi}
j_{\ell}( |{\bf k}+\alpha{\bf k}_{0}| D ) Y_{\ell m}( {{\bf k}+\alpha{\bf k}_{0} \over |{\bf k}+\alpha{\bf k}_{0}|} )  \nonumber\\
& \approx{1 \over k_{0}D } \sqrt{3 \over 4\pi} j_{\ell}(kD)Y_{\ell m}(\hat {\bf k}) \nonumber\\
&\quad - \alpha R^{1,\ell+1}_{\ell m}j_{\ell+1}(kD)Y_{\ell+1,m}(\hat {\bf k}) \nonumber\\
& \quad + \alpha R^{1,\ell-1}_{\ell m}j_{\ell-1}(kD) Y_{\ell-1,m}(\hat {\bf k})\,,
\end{align}
where the coupling matrix $R_{\ell m}^{\ell_1 , \ell_2}$ is defined as \footnote{The normalization differs 
from  Ref. \cite{wayne} by $w_\ell$.}
\begin{align} \label{coupling_matrix}
R_{\ell m}^{\ell_1 , \ell_2} =& (-1)^{m} \sqrt{ (2\ell+1)(2\ell_1+1)(2\ell_2+1) \over 4\pi}
\nonumber\\
& \times \wj{\ell_1}{\ell_2}{\ell}{0}{0}{0}  \wj{\ell_1}{\ell_2}{\ell}{0}{m}{-m} \,.
\end{align}
With these relations, the first order term in $h$ becomes 
\begin{equation}\label{temperature_dipole}
\langle T_{\ell m}^{*} T_{\ell^{\prime} m^{\prime}}\rangle^{(1)} =\delta_{mm^{\prime}} w_1
\left[  R^{1,\ell}_{\ell^{\prime}m}C_{\ell}^{TT} +  R^{1,\ell^{\prime}}_{\ell m}C_{\ell^{\prime}}^{TT} \right] \,.
\end{equation}
By virtue of the
Wigner 3$j$ symbol in the coupling matrix,
the dipole modulation couples $\ell$ to $\ell\pm 1$ with a strength that is linear in $h$
or $w_{1}$.

Likewise, the same operations on the term quadratic in $h$ yield
\begin{equation}
\langle T_{\ell m}^{*} T_{\ell^{\prime} m^{\prime}}\rangle^{(2)}
=\delta_{mm^{\prime}}  w_1^2 \left[  \sum_{\ell_{1}} R^{1,\ell_{1}}_{\ell m}R^{1,\ell_{1}}_{\ell^{\prime}m} C_{\ell_{1}}^{TT}
\right] \,.
\end{equation}
The quadratic term in $h$ or $w_{1}$ couples $\ell$ to $\ell \pm 2$.
These results are the same as obtained in Ref. \cite{wayne}.

\subsection{Modulated Polarization Field} \label{sec :general_polarization}

For predicting polarization statistics, the direct angular modulation approach of
\cite{wayne} does not apply given that the sources of
the polarization field are modulated and not the field itself.

The quadrupole source comes from the temperature field through Eq.~(\ref{quadsource})
\begin{equation}
T_{2m}({\bf x}) =- \int d\hat {\bf n}^{\prime} Y_{2m}^{*}(\hat{\bf n}^{\prime}){\Phi({\bf x}^{\prime}) \over 3}\,,
\end{equation}
where recall ${\bf x}^{\prime} = {\bf x} + \Delta D \hat{\bf n}^{\prime}$ (see Fig.~\ref{pol}).

Decomposing $\Phi$ into Fourier modes yields
\begin{eqnarray}
\!\!\!\!\! T_{2 m}({\bf x}) = \int {d^{3} k \over (2\pi)^{3}} {\Phi({\bf k}) \over 3} 4\pi  j_{2}(k \Delta D)
Y_{2 m}^{*}(\hat {\bf k}) e^{i{\bf k} \cdot {\bf x}}.
\end{eqnarray}

The Stokes parameters of the polarization are given by Eq.~(\ref{qplusiumain})
\begin{eqnarray} \label{qplusiu}
&& (Q \pm i U)(\hat{{\bf n}}) \\
&& \quad =-\frac{\sqrt{6}}{10} \int dD {d\tau \over dD} e^{-\tau}
 \sum_{m}T_{2m}(D  \hat{\bf{n}} ){}_{\pm2}Y_{2m}(\hat{\bf{n}}) \nonumber\\
&& \quad =
-\frac{\sqrt{6}}{10} \int dD {d\tau \over dD} e^{-\tau}
 {1 \over 3} \int {d^{3} k \over (2\pi)^{3}} \Phi({\bf k}) (4\pi)^{2} \nonumber\\
&& \qquad \times  j_{2}(k \Delta D)\sum_{\ell^{\prime}} i^{\ell^{\prime}} j_{\ell^{\prime}}(kD) \nonumber\\
 &&\qquad \times \sum_{m_{2}m^{\prime}} Y_{2 m_{2}}^{*}(\hat {\bf k})
Y_{\ell^{\prime}m^{\prime}}^{*}(\hat{\bf k}) Y_{\ell^{\prime}m^{\prime}}(\hat{\bf n})
{}_{\pm2}Y_{2m_{2}}(\hat{\bf{n}}) \,. \nonumber
\end{eqnarray}
Recoupling the last line of Eq. (\ref{qplusiu}), we have
 \begin{eqnarray}
 && \sum_{m_{2}m^{\prime}} Y_{2 m_2}^{*}(\hat {\bf k})
Y_{\ell^{\prime}m^{\prime}}^{*}(\hat{\bf k}) Y_{\ell^{\prime}m^{\prime}}(\hat{\bf n})
{}_{\pm2}Y_{2m_{2}}(\hat{\bf{n}})
= \nonumber\\
&&\quad  \sum_{\ell m}
{ (2\ell^{\prime}+1)5 \over 4\pi}
\wj{\ell^{\prime}}{2}{\ell}{0}{\mp 2}{\pm 2}
\wj{\ell^{\prime}}{2}{\ell}{0}{0}{0} \nonumber\\
&&\quad \times{}_{\pm2}Y_{\ell m}(\hat{\bf{n}})Y_{\ell m}^{*}(\hat {\bf k}) \,.
\end{eqnarray}

Decomposing into the spin-2 spherical harmonics, 
\begin{eqnarray}
&& E_{\ell m} \pm i B_{\ell m} 
 = -\int d\hat{\bf n}   {}_{\pm2}Y_{\ell m}^{*}(\hat{\bf{n}})(Q\pm i U)(\hat{\bf{n}})
\nonumber\\
&& \quad=
\frac{\sqrt{6}}{10} \int dD {d\tau \over dD} e^{-\tau}
 {1 \over 3} \int {d^{3} k \over (2\pi)^{3}} \Phi({\bf k}) \nonumber\\
&&\qquad \times  (4\pi)^{2} j_{2}(k \Delta D) \sum_{\ell^{\prime}} i^{\ell^{\prime}} j_{\ell^{\prime}}(kD)\\
&& \qquad \times
{(2\ell^{\prime}+1)5 \over 4\pi}
\wj{\ell^{\prime}}{2}{\ell}{0}{\mp 2}{\pm 2}
\wj{\ell^{\prime}}{2}{\ell}{0}{0}{0}  Y_{\ell m}^{*}(\hat {\bf k}) \,. \nonumber
 \end{eqnarray}
Note that the $B$-modes vanish since the second Wigner $3j$ 
symbol is zero if $(\ell+\ell^{\prime} + 2 )$ is odd and
\begin{equation}
\wj{\ell^{\prime}}{2}{\ell}{0}{2}{-2}  =(-1)^{\ell+\ell^{\prime}} \wj{\ell^{\prime}}{2}{\ell}{0}{-2}{2} \,.
\end{equation}
This is to be expected since all  scalar sources, even products of fundamental
scalars, generate only $E$-modes.

Paralleling the modulated temperature derivation of \S \ref{sec: modulated_temp_statistics},
we find that there is the zeroth order piece coming from the unmodulated field that
reproduces the standard result \cite{huandwhite}. 

\begin{eqnarray} \label{standard_polarization_form}
\langle E_{\ell m}^{*} E_{\ell^{\prime} m^{\prime}} \rangle^{(0)} = \delta_{\ell \ell^{\prime}} \delta_{m m^{\prime}} C_{\ell}^{EE} \,,
\end{eqnarray}
where
\begin{equation}
 \label{standard_polarization}
C_\ell^{EE} = {1 \over 6} \int {dk \over k} {k^{3} P_{g} \over 2\pi^{2}} 4\pi I_{\ell}^{2}(k) \,,
\end{equation}
with
\begin{equation}\label{Il}
I_{\ell}(k) = \int dD {d \tau \over dD}e^{-\tau} j_{2}(k\Delta D) \epsilon_{\ell}^{(0)}(kD) \,,
\end{equation}
and the polarization radial function
\begin{equation}
\epsilon_{\ell}^{(0)}(x) = \sqrt{{ 3 \over 8} {(\ell+2)! \over (\ell-2)!}} {j_{\ell}(x) \over x^{2}} \,.
\end{equation}

The linear piece in $h$ of the dipole modulation becomes
\begin{eqnarray} 
\label{elmelmpfirstorder}
&& \langle E_{\ell m}^{*} E_{\ell^{\prime} m^{\prime}} \rangle^{(1)} = { (4 \pi)^{2} \over 6}
 {w_{1} \over 2}\sqrt{3 \over 4\pi}  {1 \over k_{0}D_{\rm rec}} \nonumber\\
&&\quad \times \int {d^{3}k \over (2 \pi)^{3}} \int {d^{3}k^{\prime} \over (2 \pi )^{3}} \left[ P_{g}(k) +P_{g}(k^{\prime})\right] \nonumber\\
&&\quad \times (2\pi)^{3} \left[ \delta ({\bf k}^{\prime}-{\bf k}-{\bf k}_{0}) -  \delta({\bf k}^{\prime}-{\bf k} + {\bf k}_{0})\right] \nonumber\\
&&\quad \times \left[ \int dD {d \tau \over dD}e^{-\tau} j_{2} (k\Delta D) \epsilon_{\ell}^{(0)}(kD) \right] \nonumber\\
&& \quad \times \left[ \int dD {d \tau \over dD}e^{-\tau} j_{2} (k^{\prime}\Delta D) \epsilon_{\ell^{\prime}}^{(0)}(k^{\prime}D) \right]
 \nonumber\\
&&\quad  \times i^{\ell^{\prime}-\ell-1}Y_{\ell m}(\hat {\bf k})Y_{\ell^{\prime} m^{\prime}}^{*}(\hat {\bf k^{\prime}}) \,.
\end{eqnarray}

Assuming, as in the temperature case, that $k/k_{0} \ll 1$,  we can again Taylor expand the spherical Bessel functions and the spherical harmonic 
which, after some algebra, reduces Eq.~(\ref{elmelmpfirstorder}) to
\begin{align} \label{resultEEcorrelation}
 \langle E_{\ell m}^{*} E_{\ell^{\prime} m^{\prime}} \rangle^{(1)} &= \delta_{m m^{\prime}} w_1 R^{1,\ell}_{\ell^{\prime} m} \Big[\sqrt{ (\ell^{\prime} + 2)! (\ell -2)! \over (\ell^{\prime}-2)! (\ell+2)!}
A_{\ell} \nonumber\\
& \quad + (\delta_{\ell,\ell^{\prime}+1} - \delta_{\ell,\ell^{\prime}-1} ) B_{\ell \ell^{\prime}}\Big]
+ ({\ell \leftrightarrow \ell^{\prime}})\,, 
\end{align}
where
\begin{eqnarray} \label{AlBllp}
A_{\ell} &= & {1 \over 6} \int {dk \over k} {k^{3} P_{g} \over 2\pi^{2}} 4\pi
I_{\ell}(k) J_{\ell}(k)\,,\nonumber\\
 B_{\ell \ell^{\prime}} &=&
 {1 \over 6}
\int {dk \over k} {k^{3} P_{g} \over 2\pi^{2}} 4\pi I_{\ell}(k) K_{\ell^{\prime}}(k) \,,
 \end{eqnarray}
and 
\begin{align}
\label{polintegral}
J_{\ell}(k)  &=
\int dD {d \tau \over dD}e^{-\tau} {D\over D_{\rm rec}}j_{2} (k\Delta D) \epsilon_{\ell}^{(0)}(kD)\,, \nonumber\\
K_{\ell}(k)  &=
\int dD {d \tau \over dD}e^{-\tau}
{\Delta D \over D_{\rm rec}} j_3 (k\Delta D) \epsilon_{\ell}^{(0)}(kD)\,,
\end{align}
with
$I_{\ell}(k)$ defined in Eq. (\ref{Il}).  Like the temperature field, the coupling of $\ell$ to
$\ell \pm 1$ is linear in $h$.  The quadratic term can be similarly calculated but is
 more cumbersome and we omit
the derivation here.

For reasonable reionization histories, 
the dominant term in Eq.~(\ref{resultEEcorrelation}) is the first term.
Comparing Eq.~(\ref{standard_polarization}) with (\ref{polintegral}) we see that  
 to order of magnitude,
 \begin{equation}
A_{\ell} \sim
{\bar D \over D_{\rm rec}}C_{\ell}^{EE} \,,
\end{equation}
where $\bar D$ is a typical distance to reionization.  Compared with the modulation
of the temperature spectrum in Eq.~(\ref{temperature_dipole}), the modulation of the polarization spectrum is suppressed
by a factor of $\bar D / D_{\rm rec}$.  This factor is due to the lower level of spatial
modulation at typical distances to reionization in the dipolar model of Eq.~(\ref{dipole_modulation}). Typically, it has a value of $\bar D / D_{\rm rec} \sim 2/3$.  

For the $TE$ two point function, the same operations yield
\begin{align} \label{resultTEcorrelation}
\langle T_{\ell m}^* E_{\ell^{\prime} m^{\prime}} \rangle^{(1)}=&  \delta_{m m^{\prime}}  w_1 R^{1,\ell}_{\ell^{\prime} m} \Big[
\sqrt{ (\ell^{\prime} + 2)! (\ell -2)! \over (\ell^{\prime}-2)! (\ell+2)!}
H_{\ell} + \nonumber\\
& +  ( \delta_{\ell,\ell^{\prime}+1} - \delta_{\ell,\ell^{\prime}-1} ) L_{\ell \ell^{\prime}}]
+ C_{\ell'}^{TE}   \Big] \,,
\end{align}
with
\begin{align}\label{HlLl}
H_{\ell} = & {1 \over \sqrt{54} } \int {dk \over k} {k^{3} P_{g} \over 2\pi^{2}}
j_{\ell}(k D_{\rm {rec}}) 4 \pi J_{\ell}(k)\,, \nonumber\\
 L_{\ell \ell^{\prime}} =&
 {1 \over \sqrt{54}} 
\int {dk \over k} {k^3 P_g \over 2\pi^{2}} j_{\ell}(k D_{\rm rec}) 4 \pi K_{\ell^{\prime}}(k)\,,
 \end{align}
where $C_\ell^{TE}$ is the usual unmodulated cross power spectrum
\begin{eqnarray}
\label{unmodte}
 C_{\ell}^{TE} = {1 \over \sqrt{54} }  \int {d k \over k} {k^{3}P_{g}(k)\over 2\pi^{2}} 4 \pi j_{\ell}(k D_{\rm rec}) I_{\ell}(k)\,,
\end{eqnarray}
and $I_{\ell}(k)$ is defined in Eq. (\ref{Il}).

\subsection{Comparison with Simulations} \label{sec: Codetests}

In this section, we test the
3D simulation method against the 
 analytic results of \S \ref{sec: modulated_temp_statistics} and \ref{sec :general_polarization}.
  The simulations are performed according to the 
 algorithm described in Section \ref{sec: simulations} with a box of size that is
 four times the diameter of the last scattering surface and a spatial resolution defined by
 the $512^{3}$ pixels.

We first test the simulations against the usual results given by
Eq.~(\ref{unmodtemp}),    (\ref{standard_polarization}), and (\ref{unmodte})
 for an unmodulated scale invariant
potential spectrum. 
The temperature power spectra averaged over multiple realizations agrees with the 
analytic results within the sample variance
\begin{equation}
\frac{\Delta C_{\ell}^{TT}}{C_{\ell}^{TT}} = \sqrt{\frac{2}{(2 \ell + 1)N}} \,,
\end{equation}
where $N$ is the number of simulated skies.

In order to predict the $E$-modes, we calculate the quadrupolar temperature sources for 
 $80$ radii along each line of sight, and in every unit vector $\hat{\bf {n}}$ corresponding to $12,288$ HEALPIX pixel centers at resolution $N_{\rm{side}}=32$. The small evolution of perturbations of the potential field between recombination and reionization
 is accounted for via the transfer function. 

We compute the  $E$ modes for different reionization histories with ionization fraction equal to $1$ out to a redshift where the
optical depth  $\tau=\{0.0279,0.055,0.089\}$ and zero beyond this redshift. In all cases, the distribution of numerical spectra are consistent with cosmic variance for $\ell \le 10$. 
 For larger multipoles there is a diminution of power that is consistent with 
the Nyquist frequency of the box, but this does not affect any of our results.

\begin{figure}[htbp]
\begin{center}
\includegraphics[width=3in]{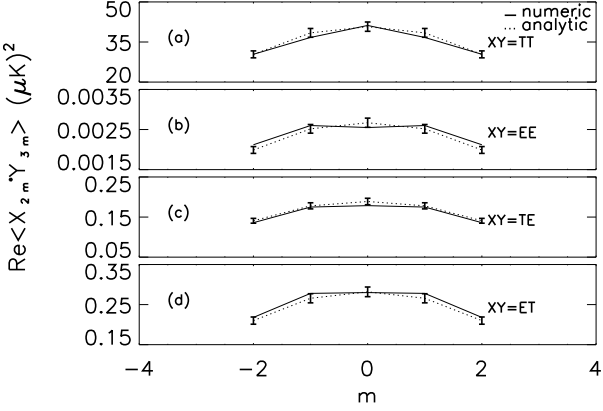}
\caption{Comparison between analytical solution (dotted line) and $545$ 3D simulations (solid line) of the dipole modulation model with $w_{1}=0.2$.  The two point correlation between 
the quadrupole and octopole $m$ moments are shown for the $X$ and $Y$ fields: $XY$ = $TT$, $EE$,
$TE$, $ET$.   Error bars reflect the sample variance expected for 545 skies.} \label{correlations_comparison}
\end{center}
\end{figure}

Finally, we check the dipole modulation model against the analytic results of the previous 
sections.
We showed there that a dipole modulation gives rise to couplings between $\ell$ and $\ell^{\prime}=\ell \pm 1$ modes in the $TT$, $EE$, $TE$ and $ET$ two point functions.
As an example of these correlations, we compare simulation results with the analytic
calculation in Fig.~\ref{correlations_comparison}.
We again
find agreement within sample variance expectations.

\subsection{Dipolar Modulation Estimators} \label{sec: Unbiased estimator}

From the analytic calculations of \S \ref{sec: modulated_temp_statistics} and \ref{sec :general_polarization}, we can construct unbiased estimators of the dipole
modulation amplitude $w_1$.  In the presence of a dipole modulation, the
expectation values of temperature and polarization modes
 $\ell$ and $\ell^{\prime}=\ell \pm 1$ are correlated.
Each pair of multipoles then forms an estimator of $w_1$
 \begin{equation}
\hat{w}_{1,\ell m}^{XY} = {X_{\ell m}^{*}Y_{\ell+1,m} \over f_{\ell}^{XY}R_{\ell+1,m} ^{1 \ell}}\,,
\end{equation} 
where $XY \in \{TT,EE,ET,TE\}$, $R_{\ell+1,m}^{1 \ell}$ are coupling matrices defined in Eq.~(\ref{coupling_matrix}) and the coefficients $f_{\ell}^{XY}$ are given by
\begin{equation}
 \left\langle X_{\ell m}^{*} Y_{\ell+1, m}\right\rangle  = f_{\ell}^{XY}R_{\ell+1,m}^{1 \ell} w_1\,,
\end{equation}
such that $\langle \hat{w}_{1,\ell m}^{XY}  \rangle =  w_1$.   Specifically,  Eqs.~(\ref{temperature_dipole}), (\ref{resultEEcorrelation}), and (\ref{resultTEcorrelation}) give
\begin{align}\label{flXY}
f_{\ell}^{TT} = & C_{\ell}^{TT} +  C_{\ell+1}^{TT}\,, \\
f_{\ell}^{EE} = & \sqrt{ (\ell + 3)! (\ell -2)! \over (\ell-1)! (\ell+2)!}A_{\ell}  -  B_{\ell,\ell+1} + ({\ell \leftrightarrow \ell+1})\,,\nonumber\\
f_{\ell}^{TE} = & \sqrt{ (\ell + 3)! (\ell -2)! \over (\ell-1)! (\ell+2)!}H_{\ell}  -  L_{\ell,\ell+1}  + C_{\ell+1}^{TE}\,, \nonumber\\
f_{\ell}^{ET} = & \sqrt{ (\ell + 2)! (\ell -1)! \over (\ell-2)! (\ell+3)!}H_{\ell+1}  +  L_{\ell+1,\ell} + C_{\ell}^{TE}\,, \nonumber
\end{align}
where $A_{\ell}$, $B_{\ell,\ell+1}$, $H_{\ell}$ and $L_{\ell,\ell+1}$ are given by Eqs. (\ref{AlBllp}) and (\ref{HlLl}).

Each estimator carries a large sample variance and so one can combine them to form the joint estimator
\begin{equation}\label{general_estimator}
\hat{w}_{1}=\sum_{XY}\sum_{\ell,m}A_{\ell m}^{XY}\hat{w}_{1,\ell m}^{XY}\,,
\end{equation}
where $A_{\ell m}^{XY}$ are the weights assigned to each one.   To maintain the condition that
the estimator is unbiased $\langle  \hat{w}_1 \rangle = w_1$, the weights must satisfy
\begin{equation}\label{constraint}
\sum_{XY}\sum_{\ell m}A_{\ell m}^{XY} = 1 \,.
\end{equation}
 Note that while
individual $\hat{w}_{1,\ell m}^{XY}$ estimators are complex, the joint estimator is real
if $+m$ and $-m$ are
combined with equal weights.

We employ  weights that minimize the variance of the joint estimator.   Given a model with $w_1=0$, the variance is
\begin{align}
\langle \hat{w}_{1}^2\rangle=
\sum_{XY,X^{\prime}Y^{\prime}} \sum_{\ell m} A_{\ell m}^{XY} A_{\ell m}^{X^{\prime}Y^{\prime}}
{\mathcal{C}_{XY,X^{\prime}Y^{\prime}}^{(\ell)} \over (R_{\ell+1,m}^{1 \ell})^{2}} \,.
\end{align}
Here $\mathcal{C}_{XY,X^{\prime}Y^{\prime}}^{(\ell)}$ are the components of the matrix $\bf {\mathcal{C}}$ given by
\begin{equation}\label{Cmatrix}
\mathcal{C}_{XY,X^{\prime}Y^{\prime}}^{(\ell)} = {C_{\ell}^{XX^{\prime}}C_{\ell+1}^{YY^{\prime}} \over f_{\ell}^{XY} f_{\ell}^{X^{\prime}Y^{\prime}}}\,,
\end{equation}
where $C_{\ell}^{XX^{\prime}}$ are the standard angular power spectra coefficients.

Imposing the normalization constraint on the weights, given by Eq.~(\ref{constraint}),
we 
minimize the variance with 
\begin{equation} \label{AlmXY}
A_{\ell m}^{XY} = 
{\sum_{X'Y'}
{[{\bf \mathcal{C}}^{(\ell )}]^{-1}_{XY,X^{\prime}Y^{\prime}} (R_{\ell+1,m}^{1 \ell})^{2}} \over
 \sum_{XY,X'Y'}  \sum_{\ell m} [{\bf \mathcal{C}}^{(\ell )}]^{-1}_{XY,X^{\prime}Y^{\prime}} (R_{\ell+1,m}^{1 \ell})^{2}}\,,
\end{equation}
which is the usual inverse covariance weight of covarying estimators.  Likewise in the case
that just one of the $XY$ combinations is used this general form reduces to
\begin{equation}\label{w1XY}
\hat{w}_{1}^{XY} = {\sum_{\ell {m}} { f_{\ell}^{XY}R_{\ell+1,m}^{1 \ell} \over C_{\ell}^{XX}C_{\ell+1}^{YY}}
 X_{\ell m}^{*} Y_{\ell+1, m}
 \over  {\sum_{\ell {m}}
  {(f_{\ell}^{XY}R_{\ell+1,m}^{1 \ell})^{2}\over C_{\ell}^{XX}C_{\ell+1}^{YY}}}}\,,
\end{equation}
which is the usual inverse
variance weighted estimator.  Note that the sum over $\ell$ can be restricted to
$\ell \le \ell_{\rm max}$ for modulation models which only affect large scales.

Finally, given instrumental noise power spectra $N_\ell^{XY}$, the variance and minimum
variance weights are modified by including the spectrum into the covariance matrix
\begin{equation}
\mathcal{C}_{XY,X^{\prime}Y^{\prime}}^{(\ell)} = {(C_{\ell}^{XX^{\prime}} + N_{\ell}^{XX^{\prime}})(C_{\ell+1}^{YY^{\prime}} + N_{\ell+1}^{YY^{\prime}}) \over f_{\ell}^{XY} f_{\ell}^{X^{\prime}Y^{\prime}}} \,,
\end{equation}
where we typically assume that the noise cross power spectrum $TE$ vanishes.

\vfill
\bibliography{paperv2}

\vfill

\end{document}